\documentclass[aps,twocolumn,longbibliography]{revtex4}

\usepackage{graphicx}
\usepackage{amssymb, amsmath,mathtools}
\usepackage{bbm}
\usepackage{enumerate}
\usepackage{bbold}
\usepackage{xcolor}
\usepackage{hyperref}

\usepackage{hyperref}
\usepackage[figure,table]{hypcap}               % to correct a problem with hyperref
\hypersetup{
pdftitle = {},
pdfsubject = {},
pdfauthor = {},
pdfkeywords = {},
pdfcreator = {},
pdfproducer = {LaTeX with hyperref},
colorlinks = true,
linkcolor = blue,
anchorcolor = blue,
citecolor = blue,
filecolor = red,
menucolor = red,
pagecolor = red,
urlcolor  = blue,
breaklinks = true,
pdfstartview = FitV,
pdfhighlight = /I,
pdfpagelayout = TwoColumn,
hypertexnames=true
}

\graphicspath{{./}{./plots/}}

\usepackage{xcolor}

\newcommand\scalemath[2]{\scalebox{#1}{\mbox{\ensuremath{\displaystyle #2}}}}

\begin{document}

\title{$p$-wave superconductivity in Luttinger semimetals}

\author{Julia M. Link and Igor F. Herbut}

\affiliation{Department of Physics, Simon Fraser University, Burnaby, British Columbia V5A 1S6, Canada}

\begin{abstract}
We consider the three-dimensional spin-orbit-coupled Luttinger semimetal of ``spin" $3/2$ particles in presence of weak attractive interaction in the $l=1$ ($p$-wave) channel, and determine the low-temperature phase diagram for both particle- and hole dopings. The phase diagram depends crucially on the sign of the chemical potential, with two different states (with total angular momentum $j=0$ and $j=3$) competing on the hole-doped side, and three (one $j=1$ and two different $j=2$) states on the particle-doped side. The ground-state condensates of Cooper pairs with the total angular momentum $j=1,2,3$ are selected by the quartic, and even sextic terms in the Ginzburg-Landau free energy. Interestingly, we find that all the $p$-wave ground states appearing in the phase diagram, while displaying different patterns of reduction of the rotational symmetry, preserve time reversal. The resulting quasiparticle spectrum is either fully gapped or with point nodes, with nodal lines being absent.
\end{abstract}

\maketitle

\section{Introduction}

Luttinger semimetals are three-dimensional materials, which due to their strong spin-orbit coupling exhibit band inversion and the concomitant parabolic dispersion at the Fermi level \cite{luttinger}.  When undoped, the vanishing density of states leaves the Coulomb interaction unscreened, and the system is expected to exhibit a non-Fermi liquid ground state \cite{abrikosov, moon, herbut1, janssen1, dora}. When the density of carriers is even slightly finite, on the other hand, many Luttinger semimetals become superconductors with sizable critical temperatures \cite{butch, bay, kim}. The resulting superconducting phases appear to be unconventional, and the pairing interaction, the pattern of broken symmetry, quasiparticle spectrum, topological characteristics, and the behavior in  the magnetic field have all been recently under investigation \cite{boettcher1, meinert, brydon, boettcher2, roy}. The main conceptual novelty arises from the fact that the effective spin of the Luttinger fermions is $3/2$. The total spin of the Cooper pair can therefore assume unusually large values, and be $s = 0,1,2,3$.  Depending on the angular momentum $l$ of the attractive pairing channel, Fermi statistics then constrains the Cooper pairs to assume quantum numbers
$(l,s)$ with $ l+s$ even. Whereas the $s$-wave state $(0,0)$ would be quite conventional and fully gapped,
the $d$-wave states $(0,2)$ \cite{boettcher1, brydon, boettcher2} and $(2,0)$ \cite{savary, venderbos}, due to their multicomponent nature already are not. In the latter case, for example, one can  show that there are two nearly degenerate BCS ground state that both break time-reversal symmetry, but have  rather different average magnetization \cite{herbut2}. Competition between various $d$-wave states in related systems has also been studied in the distant \cite{mermin, sauls} and recent past \cite{link1, mandal1}.
\begin{figure}
 \includegraphics[width=\columnwidth]{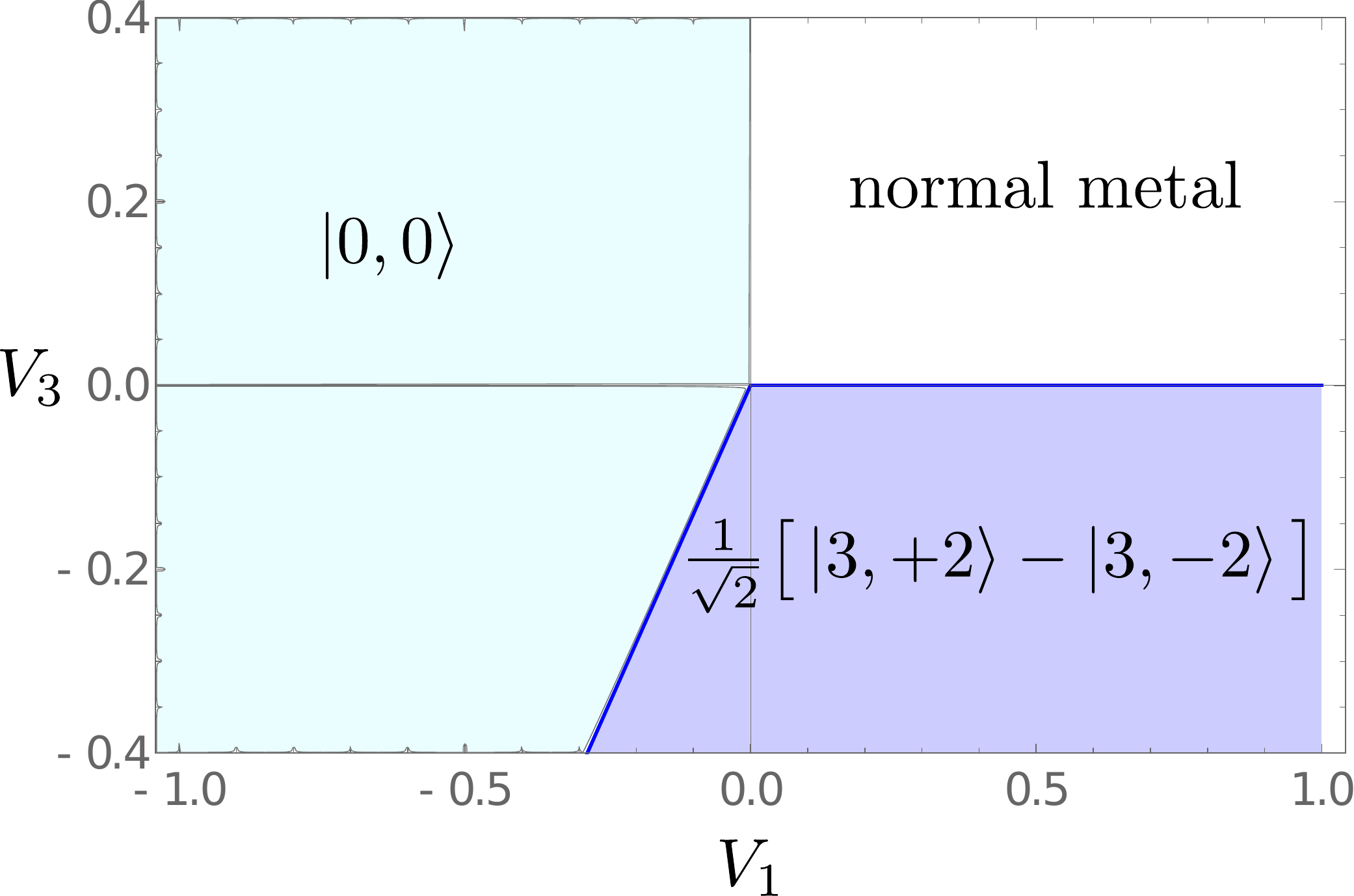}
 \caption{The low-temperature phase diagram for negative chemical potential, when the Fermi level intersects the states with the magnetic quantum number $ \pm 3/2$. There is a phase transition between the $\left|0,0 \right>$ state, which exhibits a fully gapped isotropic quasiparticle spectrum,  and the cubic-symmetric, real state $(\left| 3,+2\right> - \left| 3,-2 \right)/\sqrt{2}$, which exhibits six point nodes.}
 \label{fig:phase32}
\end{figure}

In this paper we study the superconductivity of Luttinger fermions with simplest non-local BCS pairing interaction, with the attraction in the channel with $l=1$. The relevant $p$-wave superconducting states then can have the Cooper pairs with spin $s$ either  $s=1$ or $s=3$, and allowing for the possibility that  the pairing interaction may be spin dependent, we parametrize the interaction in these two channels with two phenomenological parameters $V_1$ and $V_3$, respectively. We assume the $SO(3)$ rotational invariance, for reasons of simplicity, but also since we expect that the effect of long-range Coulomb interaction is in general to make the parabolic dispersion isotropic at low dopings \cite{abrikosov, boettcher3}. Total angular momentum of the Cooper pairs with $(l,s )=(1,1)$ can therefore be $j =0,1,2$, and with $(l,s )= (1,3)$, $j= 2,3,4$. In the weak-coupling approach two immediate questions then emerge: (1) what is the value of $j$ of the superconducting ground state at a given $V_1$ and $V_3$, and then, (2) for that value of $j$, which state in the $2j+1$-dimensional Hilbert space is the actual ground state? The answer, surprisingly, strongly depends on the sign of the chemical potential, as already noted in Ref. \cite{savary}; for $\mu < 0$ (hole doping) the electrons at the Fermi level have the magnetic quantum number $\pm 3/2$, and the computation of the superconducting susceptibility of the normal state in different superconducting channels shows that the competition is between the rotationally invariant $j=0$ state, and the $j=3$ states, with the phase boundary between the two at $V_3 / V_1 = 7/5$. Computing the coefficients of the four independent fourth-order terms in the Ginzburg-Landau (GL) free energy \cite{kawaguchi} shows then that among the $j=3$ states the lowest free energy right below the transition temperature belongs to the time-reversal-invariant state $(|3, +2\rangle - |3, -2\rangle)/\sqrt{2}$ (Fig.~\ref{fig:phase32}), in the standard notation $|j, m_{j } \rangle$, with $m_j = -j, ... j$. The phase diagram of the $j=3$ superfluid is quite intricate \cite{kawaguchi} and exhibits eleven different phases. The ground state that we obtained is symmetric under the cubic group, which happens to be the largest discrete subgroup of the $SO(3)$ that is available in its seven-dimensional irreducible representation. The $j=3$ ground state exhibits six point nodes in the quasiparticle spectrum.
\begin{figure}
 \includegraphics[width=\columnwidth]{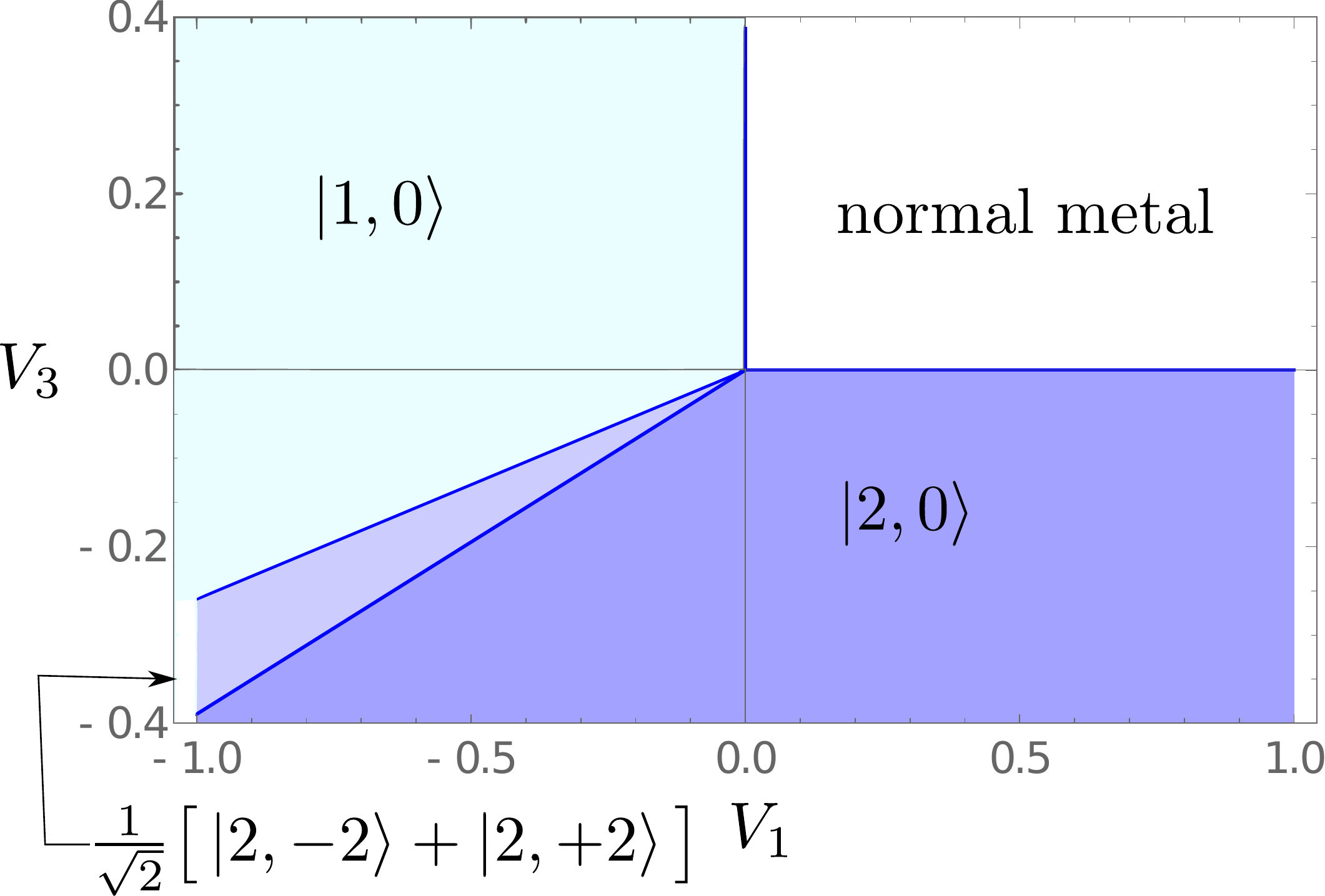}
 \caption{The low-temperature phase diagram for positive chemical potential, when the Fermi level intersects the states with the magnetic quantum number $ \pm 1/2$. The two possible values of the total angular momentum of the condensate $j$ is either one or two. If the interaction parameters $V_1$ and $V_3$ are such that the $j=1$ condensate forms, the superconducting state is the time-reversal-invariant state $\left|1,0\right>$. When the condensate has $j=2$, two superconducting ground state are possible: the uniaxial nematic state, $\left|2,0\right>$, and the biaxial nematic state, $(\left|2,-2\right> +\left|2,+2\right>)/\sqrt{2}$.}
 \label{fig:phase12}
\end{figure}

When the system is particle doped and the chemical potential is positive the Fermi level intersects the single-particle states with the magnetic quantum number $\pm 1/2$. The main competition then turns out to be between the $j=1$ and the $j=2$ states. The interesting new element is that when the interaction is attractive in both $s=1$ and $s=3$ channels, and the parameters $V_1$ and $V_3$ are both negative, $j=2$ state becomes a superposition of $(1,1)$ and $(1,3)$ pairing states \cite{savary, yu}. From the susceptibility of the normal state one finds that the $T=0$ phase transition between $j=1$ and $j=2$ states is at $V_3/V_1 = 7/27$. Considering the two independent fourth-order terms in the GL free energy, we find that the lowest energy $j=1$ state is the time-reversal-invariant state $|1,0\rangle$, which breaks the $SO(3)$  rotational symmetry down to $SO(2)$, with the quasiparticle spectrum showing two point nodes located at the axis of symmetry. On the $j=2$ side of the transition the superposition depends on the ratio $V_3/V_1$, and therefore the coefficients of the three independent fourth-order terms that determine the lowest energy state \cite{boettcher2, mermin} become functions of this ratio as well. The detailed computation of the coefficients of the fourth-order terms implies however that the ground state always preserves the time-reversal symmetry, but leaves the well-known degeneracy of such ``real" states, which is resolved only by taking the sixth-order terms in the GL free energy into the account. This finally yields the phase diagram in Fig.~\ref{fig:phase12}, where two time-reversal-symmetric states emerge: the uniaxal $SO(2)$-symmetric state $|2,0\rangle$ with the full but anisotropic gap, and the biaxial, $D_4$-symmetric state $(|2,2 \rangle+ |2,-2 \rangle)/\sqrt{2} $, with the point nodes along the $D_4$ axis.

The paper is organized as follows. In the sec.~\ref{sec:Kohn-Lutt-Ham} we define the Kohn-Luttinger Hamiltonian. In sec.~\ref{sec:p-wave-pairing} the pairing interaction in the $p$-wave channel is introduced, and in sec.~\ref{sec:GLfree-energy} the GL free energy for $j=0,1,2,3$ is presented.  In sec.~\ref{sec:oneloop} we set up the one-loop computation of the GL coefficients, and present the results on the second-order terms in sec.~\ref{sec:2ndorderGL}. The main calculation of the fourth-order terms for $j=1,2,3$ and the sixth-order terms for $j=2$ is given in the sec.~\ref{sec:4thorderGL}. Section~\ref{sec:sumdis} is the summary and brief discussion of the main results. Calculational details are presented in the Appendices.

\begin{figure}
 \includegraphics[width=\columnwidth]{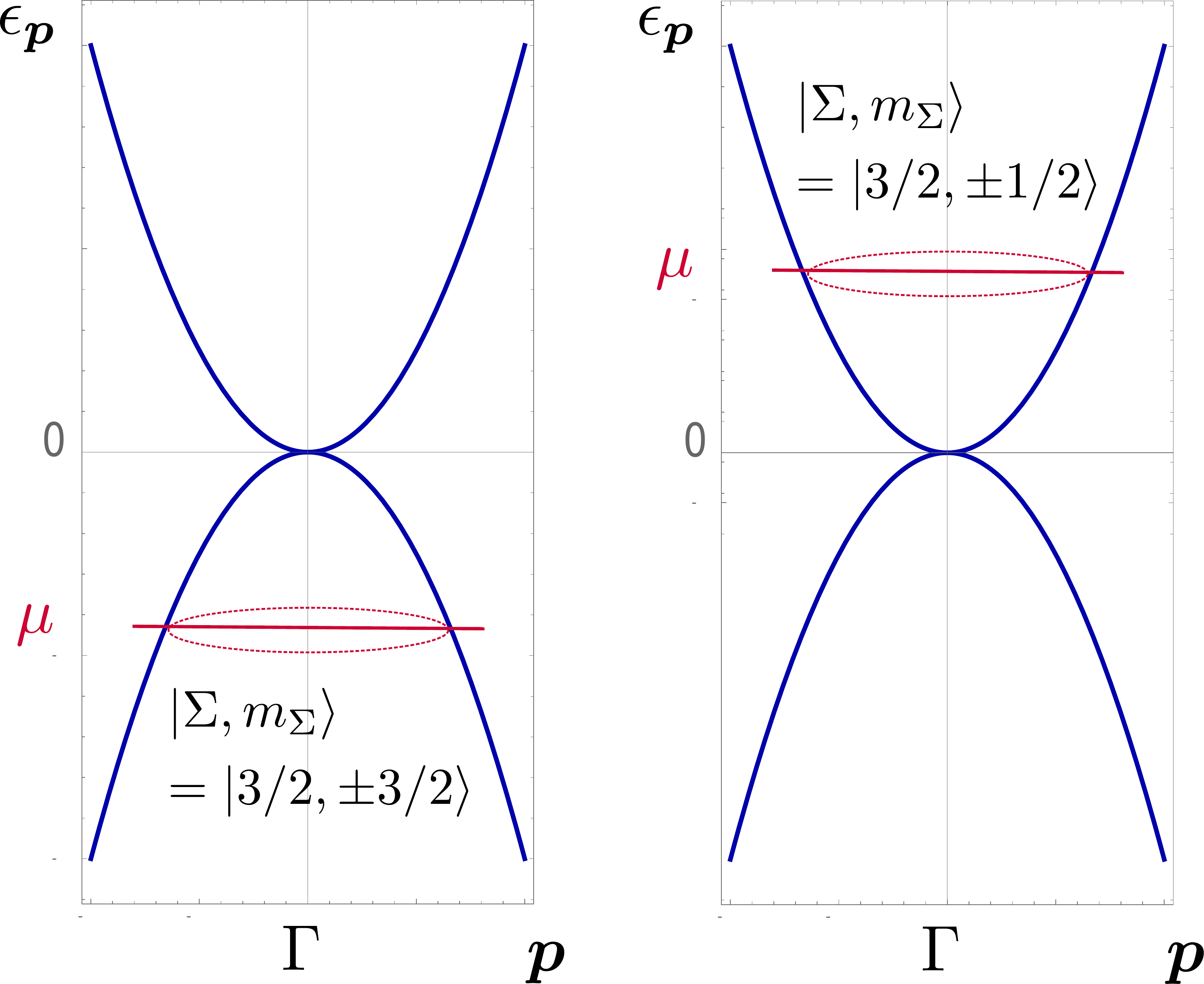}
 \caption{The energy dispersion of the Kohn-Luttinger Hamiltonian. The states at the Fermi level are either with the magnetic quantum number $\pm 1/2$ or with $\pm 3/2$, depending on the sign of the chemical potential.}
 \label{fig:energy_dispersion}
\end{figure}

\section{Kohn-Luttinger Hamiltonian}
\label{sec:Kohn-Lutt-Ham}

The single-particle Hamiltonian for the electrons in the normal state is given by the Kohn-Luttinger Hamiltonian \cite{luttinger}, which exhibits doubly (Kramers) degenerate, parabolic energy bands that touch each other at the $\Gamma$ point of the Brillouin zone:
\begin{equation}
 H_0= \frac{1}{2m} \left(\frac{5}{4} p^2 - (\boldsymbol{p}\cdot \boldsymbol{\Sigma})^2 \right) + \frac{p^2}{2 m_0}-\frac{p_a^2 \Sigma_a^2}{2 m_1} -\mu
 \:,
\end{equation}
where the summation over index $a$ runs over $a=x,y,z$, $\mu$ is the chemical potential, and the three matrices $\boldsymbol{\Sigma}=\left( \Sigma_x, \Sigma_y, \Sigma_z \right)$ is the spin--$3/2$ representation of the Lie algebra of $SO(3)$. (The explicit form of the $\Sigma$-matrices can be found in the Appendix~\ref{app:LuttAndGreen}.) The first term of the Hamiltonian is particle-hole and $SO(3)$ rotationally symmetric. The second term breaks the  particle-hole symmetry, and the third term reduces the $SO(3)$ symmetry down to the discrete cubic symmetry. We will assume a particle-hole and fully rotationally symmetric dispersion, and set $m_0 = m_1 =\infty$. We will also set the mass $m=1/2$ hereafter, for simplicity. The Luttinger Hamiltonian also has the time-reversal and inversion symmetries. We suppress the terms, which are of the first or the third order in $\boldsymbol{\Sigma}$ that would break inversion symmetry, and which are believed to be small perturbation \cite{kim}.

The energy dispersion is shown in Fig.~\ref{fig:energy_dispersion}. When the chemical potential lies in the particle band with $\mu>0$, i.e., the system is electron doped, the states crossing the Fermi level have the magnetic quantum number $\pm 1/2$. If the system is hole doped and $\mu<0$, the states at the Fermi level are with the magnetic quantum number  $\pm 3/2$. The nature of the single-particle states that are being paired will become important for the superconducting phase diagram, as was already shown by Savary \textit{et al.} \cite{savary}.

\section{$p$-wave pairing}
\label{sec:p-wave-pairing}

We define next the general rotationally invariant Hamiltonian with the BCS pairing interaction between Luttinger (spin=$3/2$) four-components fermions
$\psi$:
\begin{widetext}
\begin{equation}
 \label{eq:lag1}
 \mathcal{H} = \sum_{\mathbf{p}} \psi^\dagger (\mathbf{p})  H_0 \psi (\mathbf{p})  +
\sum_{  \mathbf{p}, \mathbf{q} } V( \theta_{ \mathbf{p}, \mathbf{q} } )
  \sum_{s=0}^3 \lambda_s \sum_{m= -s}^s  \big[ \psi^\dagger(\mathbf{p}) S_{s,m} U \psi^*(-\mathbf{p}) \big] \big[ \psi^{\rm T}(\mathbf{q}) U^\dagger S_{s,m} \psi(-\mathbf{q}) \big]
  \:,
 \end{equation}
\end{widetext}
where $U$ is the unitary part of the time-reversal operator $T$, and its explicit form can be found in Appendix~\ref{app:LuttAndGreen}. For a fixed spin quantum number $s$, $S_{s,m}$ are the $(2s+1)$ four-dimensional Hermitian matrices, which under $SO(3)$ transform as an irreducible tensor of rank $s$.  $\lambda_s$ are coupling constants. (The precise form of the matrices $S_{s,m}$ can be found in Appendix~\ref{app:interations}.) $\theta_{ \mathbf{p}, \mathbf{q} }$ is the angle between the momenta $\mathbf{p}$ and $\mathbf{q}$, and the interaction can be decomposed as
\begin{equation}
V(\theta) = \sum_{l=0}^\infty v_l P_l (\cos \theta),
\end{equation}
where $P_l (x)$ are the Legendre polynomials. The interaction is symmetric under $SO_L (3)\times SO_S (3)$, i.e., under separate rotations in the orbital and in the spin space.

The Fermi statistics implies that for even $l$ only the terms with $s=0$ and $s=2$ in the above sum are finite. The $l=0$ and $l=2$ channels were studied earlier \cite{boettcher2, herbut2, yu}. In this paper we are interested in  terms with odd $l$, and in particular with $l=1$. Only the terms with the first rank ($s=1$) and third rank ($s=3$) tensors  then contribute.  One can then rewrite the Hamiltonian in terms of the pairing matrices that describe the total angular momentum of the Cooper pair:
\begin{widetext}
\begin{equation}
 \label{eq:lag2}
 \mathcal{H} =  \sum_{\mathbf{p}}  \psi^\dagger (\mathbf{p}) H_0  \psi (\mathbf{p}) +
\sum_{  \mathbf{p}, \mathbf{q} } \sum_{s=1,3} \sum_{j=s-1}^{s+1} \sum_{m_j =-j}^j
V_s \sum_{  \mathbf{p} } \big[ \psi^\dagger(\mathbf{p}) J_{j,m_j}^{(s)}(\mathbf{p}) U \psi^*(-\mathbf{p}) \big]
\sum_{ \mathbf{q} } \big[ \psi^{\rm T}(\mathbf{q}) U^\dagger J_{j,m_j}^{(s)} (\mathbf{q} ) \psi(-\mathbf{q}) \big]
\:,
\end{equation}
\end{widetext}
where $V_s = v_1 \lambda_s$, and $J_{j,m_j}^{(s)}(\mathbf{p})$ denotes the pairing matrix of the total angular momentum $j$ which consists of the orbital angular momentum $l=1$ and the spin $s$. The pairing matrix is defined as
\begin{widetext}
 \begin{equation}
  J^{(s)} _{j,m_j} (\textbf{p})=
  \sum_{m_1 + m_s = m_j} \langle l=1, m_1, s, m_s | j, m_j \rangle Y_{1,m_1}(\textbf{p}) S_{s, m_s} ,
 \end{equation}
 \end{widetext}
where $\langle l=1, m_1,s, m_s | j, m_j \rangle $ are the standard Clebsch-Gordan coefficients and $Y_{1,m_1} (\textbf{p}) $ are the three $l=1$ spherical harmonics. We choose the normalization of the pairing matrices so that when the angular integration is performed, we find
 \begin{equation}
  \int d \hat{\mathbf{p}} \mathrm{Tr}\big[J^{(s)}_{j, m_j}  (\mathbf{p}) J^{(s) \dagger}_{j, m_j} (\mathbf{p}) \big] =\frac{4 \pi}{3},
 \end{equation}
where no summation over the indices is implied and $\int d \hat{\mathbf{p}} = \int_0^{2\pi} d\varphi \int_0^{\pi}\sin(\tau) d\tau$
 \footnote{This choice of normalization introduces an additional prefactor of $1/\sqrt{5}$ in the $J^{(1)}_{j,m_j}$ pairing channel. In other words: the Clebsch-Gordan coefficients of the $J^{(1)}_{j,m_j}$ pairing channel are multiplied by the factor $1/\sqrt{5}$.}. The explicit form of the spherical harmonics and the pairing matrices can be found in Appendix~\ref{app:interations}.

\section{Ginzburg-Landau free energy}
\label{sec:GLfree-energy}

The interaction term introduced in Hamiltonian~\eqref{eq:lag2} leads to the development of the following order parameters:
\begin{equation}
 \Delta_{j,m_j}^{(s)} =\sum_{ \mathbf{q} } \left< \psi^{\rm T}(\mathbf{q}) U^\dagger J_{j,m_j}^{(s)} (\mathbf{q} ) \psi(-\mathbf{q})  \right>
 \:.
\end{equation}
$\Delta_{j,m_j}^{(s)}$ form a complete set of $p$-wave superconducting orders. Although the pairing interaction has the enlarged $SO_L (3)\times SO_S (3)$ symmetry, the kinetic energy (Luttinger) term has only the $SO_{L+S} (3)$ symmetry, and therefore the order parameters with different values of total angular momentum $j$ do not mix in the ordered phase. We can therefore study the GL free energy for each value of the angular momentum $j$ separately.
The expansion of the GL free energy for the order parameter with given $j$ has the form:
\begin{equation}
 F^{j}(\Delta)= F_2^{j}(\Delta) + F_4^{j}(\Delta)+ F_6^{j}(\Delta) + \mathcal{O}(\Delta^8)
 \:,
\end{equation}
where the quadratic term is defined as
\begin{eqnarray}
 F_2^{j}(\Delta)&=& r^{(s_1 s_2)}_j \sum_{m_j} \Delta_{j,m_j}^{(s_1)*} \Delta_{j,m_j}^{(s_2)}
 \:.
\end{eqnarray}
In general the coefficient $r^{(s_1 s_2)}_j$ is a single number, except for $j=2$ when it becomes a two-dimensional matrix. In the later case we diagonalize this ``mixing matrix" and monitor the lower eigenvalue as a function of temperature.  The winning superconducting order sets in at the highest  temperature at which some quadratic coefficient, including the lower eigenvalue for $j=2$, $r^{(s_1 s_2)}_j$ becomes negative. This determines the total angular momentum of the ground state $j=j(V_1, V_2, \mu)$.

For each specific $j$ of the Cooper pair, there is $2j+1$-dimensional Hilbert space of states with different residual symmetries competing for the ground state. To determine the ultimate superconducting state one needs to study the higher-order terms in the GL free energy. The structure of these and even their number, however, depends on the value of $j$.  The quartic terms may be written as \cite{kawaguchi}
 \begin{widetext}
 \begin{eqnarray}
  F^{j=1}_4 &=& \lambda_1 \left| \left< \Delta | \Delta \right> \right|^2 + \lambda_2  \left< \Delta| \vec{j}_{=1} | \Delta \right> ^2 \:, \label{eq:f41} \\
  F^{j=2}_4 &=& q_1 \left| \left< \Delta | \Delta \right> \right|^2 + q_2  \left< \Delta| \vec{j}_{=2} | \Delta \right> ^2  +q_3 \left| \right< \Delta| \mathcal{T}| \Delta \left> \right|^2 \:, \label{eq:f42}\\
  F^{j=3}_4 &=& c_1 |\left< \Delta | \Delta \right>|^2 + c_2 \left< \Delta | \vec{j}_{=3} | \Delta \right> ^2 + c_3 \left| \right< \Delta| \mathcal{T}| \Delta \left> \right|^2 + c_4 \sum_{M_J} | \left< \Delta |\mathcal{T} T^{M_J}_{J=2} | \Delta \right>|^2 \label{eq:f43}
  \:,
 \end{eqnarray}
 \end{widetext}
with
 \begin{eqnarray}
   \left< \Delta | \Delta \right>  &=& \sum_{m_j} \Delta^*_{j,m_j} \Delta_{j,m_j} \:,\\
  \left< \Delta| \vec{j} | \Delta \right> &=& \sum_{m_j m_{\tilde{j}}}
  \Delta^*_{j,m_j} \vec{j}_{m_j m_{\tilde{j}}} \Delta_{j, m_{\tilde{j}}} \:, \\
  \left< \Delta| \mathcal{T} | \Delta \right> &=&
  \sum_{m_j m_{\tilde{j}}}
  \left< j, m_{j}, j,m_{\tilde{j}} | 0,0 \right>
  \Delta_{j,m_j}  \Delta_{j, m_{\tilde{j}}} \:, \\
  \left< \Delta |\mathcal{T} T^{M_j}_{J=2} | \Delta \right> &=&
  \sum_{m_j m_{\tilde{j}}}
  \left< j, m_{j}, j,m_{\tilde{j}} | 2,M_J \right>
  \Delta_{j,m_j}  \Delta_{j, m_{\tilde{j}}}
  \:.
 \end{eqnarray}
 The coefficients $a_1$ with $a \in \{ \lambda,q,c \}$ multiply the square of the absolute value of the norm of the superconducting condensate. The coefficients $a_1$ are positive, so that the free energy is bounded from bellow. The signs of other coefficients can vary. The coefficients $a_2$ multiply the square of the average magnetization of the superconducting state: if  $a_2<0$, the coexistence of the superconductivity and magnetization is preferred, whereas if $a_2 > 0$ it is not. Similarly, the term with $a_3$ that appears when $j\geq 2$ governs the preference for time-reversal symmetry breaking of the superconducting ground state: if $a_3<0$, a ``real" state which preserves time-reversal symmetry is preferred, whereas for $a_3>0$ time-reversal symmetry breaking is advantageous. Note that for $j\geq 2$ time-reversal symmetry breaking is not synonymous with magnetization, and states which are orthogonal to their time-reversed copies but nevertheless have zero average magnetization also exist \cite{mermin, kawaguchi, boettcher2}.
 The term that multiplies $c_4$ describes the ``nematicity" of the $j=3$ state \cite{kawaguchi}. The signs and magnitudes of the coefficients $a_{2,3,4}$ together determine the superconducting ground state of the condensate with the total angular momentum $j$. The phase diagram for $j=2$  in terms of the quartic coefficients of the GL free energy was first obtained by Mermin \cite{mermin, boettcher2}, and for $j=3$ by Kawaguchi and Ueda \cite{kawaguchi}.

\section{One-loop computation}
\label{sec:oneloop}
 The GL free energy is obtained by integrating out the fermionic degrees of freedom for a constant superconducting-order parameter, and then by expanding the resulting expression in powers of the order parameter to the fourth (or, if necessary, the sixth) order. At the second order in expansion the following expression is found:
 \begin{equation}
  r^{( s_1 s_2 )}_j= \frac{\delta_{ab} \delta_{s_1 s_2 }}{|V_{s_1} |}- 2 K_{j, ab}^{(s_1 s_2 )},
  \label{eq:quadraticGLorder}
 \end{equation}
where we consider the case of attractive interaction ${V_s <0}$.  $K_{j, ab}^{(s_1 s_2 )}$ is given by the expression
 \begin{equation}
  K_{j, ab}^{(s_1 s_2 )}= \mathrm{Tr} \int_Q^{\Lambda} G_0(-\omega,-\textbf{p}) J_{j,m_a}^{(s_1 )  \dagger} (\mathbf{p}) G_0(\omega,\textbf{p}) J_{j,m_b}^{(s_2 )} (\mathbf{p})
  \:,
    \label{eq:secondOrderKab}
 \end{equation}
 where $G_0(\omega, \textbf{p})=(i\omega-H_0 )^{-1}$ is the Green's function with the fermionic Matsubara frequency $\omega=(2n+1)\pi T$ and the temperature $T$. Its exact form can be found in Appendix~\ref{app:LuttAndGreen}. The measure of the integral is given by
 \begin{equation}
  \int_Q^{\Lambda}:=T \sum_{n \in \mathbb{Z}} \int_{\textbf{p}}^{\Lambda} := T \sum_{n \in \mathbb{Z}} \int_{p\leq \Lambda} \frac{d^3 p}{(2 \pi)^3}
  \:,
 \end{equation}
with the ultraviolet cutoff $\Lambda \gg \mu,T$.

Similarly, the one-loop integral that defines the quartic order of the Ginzburg-Landau free energy is given by
 \begin{equation}
 F_4^{j }(\Delta)= 4  K_{abcd}^{(s_1 s_2 s_3 s_4 )} \Delta_{j,a}^{(s_1)*} \Delta_{j,b}^{(s_2)} \Delta_{j,c}^{(s_3)*} \Delta_{j,d}^{(s_4)},
 \label{eq:quarticorderGL}
 \end{equation}
with
\begin{widetext}
 \begin{equation}
  K_{abcd}^{(s_1 s_2 s_3 s_4 )}= \mathrm{Tr} \int_Q^{\Lambda} G_0(-\omega,-\textbf{p}) J_{j,m_a}^{(s_1)\dagger} (\mathbf{p}) G_0(\omega,\textbf{p})  J_{j,m_b}^{(s_2)}   (\mathbf{p})
   G_0(-\omega,-\textbf{p}) J_{j,m_c}^{(s_3) \dagger} (\mathbf{p}) G_0(\omega,\textbf{p}) J_{j,m_d}^{(s_4)} (\mathbf{p})
  \:.
 \end{equation}
 \end{widetext}
 By inserting different superconducting states in Eq.~\eqref{eq:quarticorderGL} we find matching conditions to extract the coefficients $a_i$ \cite{boettcher2, link1} (see also Appendices \ref{ap:leading_order_coeff} and \ref{ap:matching_conditions}).

\section{Second-order terms of GL free energy}
 \label{sec:2ndorderGL}
In this section we determine the coefficients of the quadratic terms of the Ginzburg-Landau free energy, $r_j^{(s_1 s_2)}$,  and the critical temperature for all possible condensates with $j=0,1,2,3,4$.

After performing the finite-temperature Matsubara sum one can expand the integrand around the Fermi surface of the normal state,  \cite{boettcher2} and find the following result for $K_{j,ab}^{(s_1 s_2 )}$ (Appendix \ref{ap:leading_order_coeff}):
 \begin{eqnarray}
  K_{j, ab}^{(s_1 s_2 )}
 &=&\frac{c_{j, ab}^{(s_1 s_2 )}}{\pi^2} \bigg( \frac{1}{V_c} + \sqrt{|\mu|} \log\frac{|\mu|}{T} \bigg)
  \:,
  \label{eq:secondOrderKabResult}
 \end{eqnarray}
where $c_{j, ab}^{(s_1 s_2 )}= c_j ^{(s_1 s_2) }  \delta_{ab} $ are numerical coefficients corresponding to different superconducting channels. One finds the standard Cooper log-divergence with temperature, with the finite value of the one-loop integral  at $\mu=0$ defining the non-universal critical interaction $V_c$. For the condensates with $j\neq 2$ when $ c_j ^{(s_1 s_2) } = c_j  \delta_{s_1 s_2 } $ the critical temperature is given by
\begin{equation}
T_{c,j} = |\mu| \exp\bigg[-\frac{ \pi^2}{2 |V_s| c_j \sqrt{|\mu|} } \bigg( 1 - \frac{c_j |V_s|}{\pi^2 V_c} \bigg) \bigg]
 \:.
 \label{eq:criticaltemp1}
\end{equation}
The order parameter with the largest value of the coefficient $c_j$ will therefore have the highest critical temperature, and will be the one that would form
below at $T<T_{c,j}$. The list of the different values of $c_j$ is shown in Table~\ref{tab:coeff}, which is in full accordance with the ref. \cite{savary}.
For the condensate with the quantum number $j$ the value of the critical temperature crucially depends on the sign of the chemical potential $\mu$. Consequently, completely different superconducting phases are found for positive and negative chemical potentials.
 \begin{table}
 \label{tab:coeff}
  \begin{tabular}{c  c  c}
  \hline
  \hline
    (l,s,j) & $c_j$ & $c_j$\\
    & $3/2$-bands & $1/2$-bands \\
\hline
   (1,1,0) & $\frac{3}{40}$ & $\frac{1}{120}$\\
   (1,1,1) & $0$ & $\frac{1}{30}$ \\
   (1,3,3) & $\frac{3}{56}$ & $\frac{1}{280}$\\
   (1,3,4) & $\frac{13}{840}$ & $\frac{9}{280}$ \\
   \hline
   \hline
  \end{tabular}
\caption{The channel-dependent coefficients $c_{j}$ that multiply Cooper-log.}
 \end{table}

In the special case of $j=2$ there exists a mixing of two different pairing channels, since both spin values $s=1$ and $s=3$ can yield a Cooper pair with $j=2$. The quadratic coefficient of the GL free energy then becomes a matrix. For the $3/2$-energy band we find:
\begin{equation}
r_{2}^{(s_1 s_2 )}=
 \begin{pmatrix}
  \frac{1}{|V_1|}- \frac{3}{50} x & + \frac{3}{50 \sqrt{14}} x\\
  +\frac{3}{50 \sqrt{14}} x & \frac{1}{|V_3|} - \frac{3}{700} x
 \end{pmatrix}
\end{equation}
with $x= (\sqrt{|\mu| }  / \pi^2 )  \log(|\mu|/T)$. For the $1/2$-energy band
\begin{equation}
r_2^{(s_1 s_2 )}=
\begin{pmatrix}
  \frac{1}{|V_1|}-\frac{7}{150} x & - \frac{11}{50 \sqrt{14}} x\\
  -\frac{11}{50 \sqrt{14}} x & \frac{1}{|V_3|} - \frac{59}{700} x
 \end{pmatrix}
\label{eq:rkl12band}
\end{equation}
where we neglected the finite part of the loop integral. The critical temperature $T_c$ for $j=2$ is determined by the highest temperature for which
$\det \big(r_2^{(s_1 s_2 )} \big)=0$.

Let us now compare the different critical temperatures for the $3/2$-energy bands, i.e., when  $\mu<0$. We find that the two channels $(1,1,0)$ and $(1,3,3)$ have the largest coefficients $c_j $ and therefore compete in the phase diagram. The phase boundary between the condensate with $j=0$ and $j=3$ is at
\begin{equation}
 V_3 = \frac{7}{5} V_1
 \:.
\end{equation}
This phase boundary can be seen in Fig.~\ref{fig:phase32}.

In the case of positive chemical potential, $\mu>0$, the two condensates with the highest critical temperatures have a total angular momentum of $j=1$ and $j=2$. The critical temperature of the $j=2$ condensate depends both on $V_1$ and $V_3$ and is given by:
\begin{widetext}
\begin{eqnarray}
 T_{c,j=2}
 = |\mu| \exp&\big[ &\frac{ \pi^2}{2|V_1| |V_3|\sqrt{\mu} } \big(98 |V_1| + 177 |V_3| - \sqrt{9604 |V_1|^2+ 26292 |V_1| |V_3| +31329 |V_3|^2} \big) \big] \nonumber \:.
\end{eqnarray}
\end{widetext}
By comparing $T_{c, j=2}$ with the critical temperature of the $j=1$ condensate defined in Eq.~\eqref{eq:criticaltemp1} we find the phase boundary between the two phases to be at
\begin{equation}
 V_3=\frac{7}{27} V_1
 \:,
\end{equation}
as can be seen in Fig.~\ref{fig:phase12}.

\section{Fourth-order terms of GL free energy}
 \label{sec:4thorderGL}
 In the previous section we found that when $\mu<0$, depending on the values of $V_1$ and $V_3$ either the condensate with the quantum number $j=0$ or with $j=3$ forms at low temperatures. In contrast, when $\mu>0$, we find either the superconducting phase with $j=1$ or with $j=2$.

 For all the superconducting phases with $j>0$ there is a $2j+1$-dimensional Hilbert space of macroscopic quantum states that compete for the minimum of the GL free energy. The winner depends strongly on quartic terms order in the GL free energy, i. e. on the coefficients $a_i$ defined in Eq.~\eqref{eq:f41}, \eqref{eq:f42}, and \eqref{eq:f43}.
\subsection{$j=1$}
The fourth-order term in the GL free energy is defined in Eq.~\eqref{eq:f41},
where the term multiplying $\lambda_1$ denotes the norm and the term multiplying $\lambda_2$ describes the average magnetization of the condensate. For $j=1$, there exist two (modulo rotations)
possible states that minimize the free energy \cite{kawaguchi};  when $\lambda_2>0$, the state
$\left|j, m_j \right> = \left|1,0\right>$ with minimal (zero) average magnetization is the minimum, whereas if $\lambda_2<0$ the  state
$\left|j, m_j \right>= \left|1,+1\right>$ with maximal (unity) average magnetization minimizes the free energy. Both states break the normal state's symmetry $SO_{L+S} (3)$ down to $ SO(2)$. The former state preserves the time-reversal symmetry, whereas the latter breaks it maximally.

We find the following one-loop expressions for the coefficients $\lambda_1$ and $\lambda_2$ (Appendix \ref{ap:leading_order_coeff}):
\begin{widetext}
\begin{equation}
 \lambda_1= T\sum_n\int_0^\Lambda d p \frac{p^2 \left(-7 p^8-16 p^6 \mu +p^4 \left(30 \mu ^2-38 \omega_n ^2\right)+80 p^2 \mu  \left(\mu ^2+\omega_n ^2\right)+41 \left(\mu ^2+\omega_n ^2\right)^2\right)}{375 \pi ^2 \left(p^8-2 p^4 \left(\mu ^2-\omega_n ^2\right)+\left(\mu ^2+\omega_n ^2\right)^2\right)^2}
 \label{eq:lambda1}
\end{equation}
and
\begin{equation}
 \lambda_2 = T \sum_n \int_0^\Lambda dp \frac{p^2 \left(53 p^8+104 p^6 \mu +p^4 \left(30 \mu ^2-38 \omega_n ^2\right)-40 p^2 \mu  \left(\mu ^2+\omega_n ^2\right)-19 \left(\mu ^2+\omega_n ^2\right)^2\right)}{750 \pi ^2 \left(p^8-2 p^4 \left(\mu ^2-\omega_n ^2\right)+\left(\mu ^2+\omega_n ^2\right)^2\right)^2}
 \:.
 \label{eq:lambda2}
\end{equation}
\end{widetext}
Since the $j=1$ states appear in the phase diagram only when the chemical potential intersects $\pm 1/2$-band, we evaluate the above one-loop integrals for $\mu > 0$. In the weak-coupling limit, to the leading order in small parameter $T_c/ \mu$ we find \cite{zwerger}
\begin{equation}
\lambda_1 = 2\lambda_2 = \frac{4}{375} \frac{ 0.10657\sqrt{\mu} }{\pi^2 T_c^2},
\end{equation}
and both coefficients therefore positive. In the portion of the phase diagram at $\mu>0$ with the $j=1$ state, the actual superconducting ground state is the
time-reversal-preserving state $\left| 1, 0\right>$. The energy spectrum of the Bogoliubov-de Gennes quasiparticles exhibits two point nodes at the axis of the residual $SO(2)$ symmetry of the state.

\subsection{$j=2$}

\begin{figure}
 \includegraphics[width=\columnwidth]{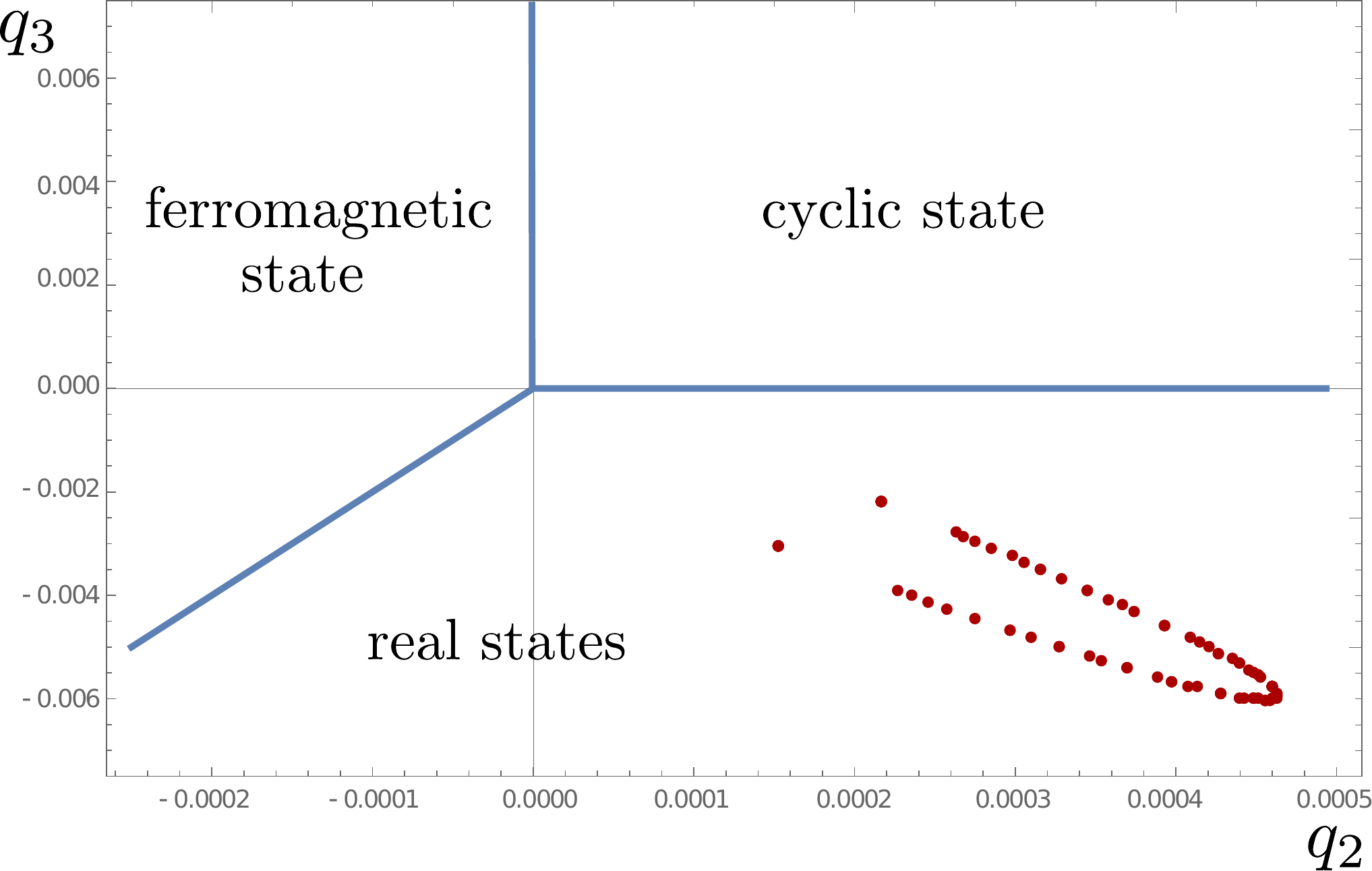}
 \caption{The phase diagram and the actual values of the coefficients of the fourth order terms in  the $j=2$ GL free energy $q_2$ and $q_3$, for various values of the interactions $V_1 <0$ and $V_3 < 7 V_1 / 27$, and for $\mu>0$. The coefficients were calculated assuming $\mu=1$ and $T_c=0.1$.}
 \label{fig:q2q3coefficients}
\end{figure}

In this section, we determine the superconducting ground state when $j=2$, which appears in the phase diagram when $\mu>0$. The quartic terms in the GL free energy are given by Eq.~\eqref{eq:f42}, and the signs and the magnitudes of the coefficients $q_2$ and $q_3$ determine the minimum of the free energy as in Fig.~\ref{fig:q2q3coefficients} \cite{mermin, link1, kawaguchi}.

The general phase diagram for $j=2$ condensate can be understood in the following way.  If $q_3>0$ and $q_2 =0$, a state that breaks time-reversal symmetry maximally, i. e., a state that is orthogonal to its time-reversed copy, is favored. There are then two candidate superconducting ground states, namely the ferromagnetic state $\left| 2,\pm 2\right>$  and the cyclic state $\frac{1}{2}(\left|2,-2\right> +\left| 2, +2\right>) + i \frac{1}{\sqrt{2}} \left| 2,0 \right>$. Since the coefficient $q_2$ multiplies the average magnetization of the state, for $q_3>0$ and $q_2< 0$ the ferromagnetic state with maximal average magnetization (two) wins, whereas for $q_3 > 0$ and $q_2>0$ the cyclic state with minimal average magnetization (of zero) wins.

Similarly, if $q_2 >0$ and $q_3 =0$, a state with minimal (zero)  average magnetization minimizes the quartic term. For small $q_3 >0$ such a state should also exhibit maximal breaking of the time-reversal, and the cyclic state therefore minimizes the free energy for all $q_2 >0$ and $q_3 >0$. For $q_3 <0$, on the other hand, the state should be invariant under time reversal, so that the term that multiplies $q_3$ is maximized. There is a multitude of such ``real" states, and to determine which of the real states is the ground state the sixth-order terms in the GL free energy need to be invoked. Finally, when both coefficients $q_3<0$ and $q_2 <0$ there is a phase transition at $q_3=20 q_2 $ between the ferromagnetic state and the (sixth-order-term-selected) real state.

We calculate the coefficients $q_{1,2,3}$ by evaluating the appropriate one-loop integrals defined in Eq.~\eqref{eq:quarticorderGL}. Special attention has to be given to the pairing matrices $J_{2,m_j}$, since when the condensate with $j=2$ is the ground state, the pairing matrices $J_{2,m_j}$ is a superposition of two different channels $(l,s,j)=(1,1,2)$ and $(l,s,j)=(1,3,2)$:
\begin{equation}
 J_{2,m_j}=z J_{2,m_j}^{(1)}+y J_{2,m_j}^{(3)}
 \:,
 \label{eq:pairingmatrixj2}
\end{equation}
where the coefficient $(z,y)^T$ is the zero eigenstate of the mixing matrix in Eq.~\eqref{eq:rkl12band} at $T=T_c$. One finds that

\begin{widetext}

 \begin{equation}
 y= \frac{66 |V_3|}{ \sqrt{  (14 |V_1| - \frac{177}{7} |V_3| ) \sqrt{9604 |V_1|^2+26292 |V_1| |V_3|+31329 |V_3|^2}
 +1372 |V_1|^2  - 600 |V_1| |V_3| + \frac{61821}{7} |V_3|^2   }  } ,
 \end{equation}

\end{widetext}
and $z= \sqrt{1-y^2}$. In the special case where $|V_1|=|V_3|$, for example, we find that ${J_{2,m_j}=0.59 J_{2,m_j}^{(1)}+ 0.81 J_{2,m_j}^{(3)}}$ \cite{savary}. Using the pairing matrices in Eq.~\eqref{eq:pairingmatrixj2}, we find  the following expressions for the coefficients $q_1$, $q_2$, and $q_3$:
\begin{widetext}
 \begin{eqnarray}
  q_1&=& T \sum_n \int_0^\Lambda dp
  \frac{p^2}{385875 \pi ^2 \left(2 \omega _n^2 \left(\mu ^2+p^4\right)+\omega _n^4+\left(p^4-\mu ^2\right)^2\right)^2}
  \big[
  2 p^4 \big(\mu ^2 (45251 y^4+42936 \sqrt{14} y^3 z+149604 y^2 z^2 \nonumber \\
  &+&
  21924 \sqrt{14} y z^3+36456 z^4)-5 \omega _n^2 (2283 y^4-6696 \sqrt{14} y^3 z+3780 y^2 z^2-3276 \sqrt{14} y z^3+392 z^4)\big) \nonumber  \\
  &+&
  8 \mu  p^2 \left(9858 y^4+7623 \sqrt{14} y^3 z+24262 y^2 z^2+6762 \sqrt{14} y z^3-1862 z^4\right) \left(\mu ^2+\omega _n^2\right) \\
  &+&
  7 \left(4393 y^4+1248 \sqrt{14} y^3 z+17122 y^2 z^2-168 \sqrt{14} y z^3+4508 z^4\right) \left(\mu ^2+\omega _n^2\right){}^2
  +
  5 p^8 \big(3335 y^4+1968 \sqrt{14} y^3 z \nonumber \\
  &+&
  11074 y^2 z^2+336 \sqrt{14} y z^3+1372 z^4\big)+8 \mu  p^6 \left(7302 y^4+5757 \sqrt{14} y^3 z+28238 y^2 z^2+3318 \sqrt{14} y z^3-98 z^4\right)
  \big] \nonumber
  \:,
 \end{eqnarray}
 \begin{eqnarray}
  q_2&=& T \sum_n \int_0^\Lambda dp
  \frac{p^2}{771750 \pi ^2 \left(2 \omega _n^2 \left(\mu ^2+p^4\right)+\omega _n^4+\left(p^4-\mu ^2\right)^2\right){}^2}
  \big[
  2 p^4 \bigg(8 y^4 \left(1337 \mu ^2+717 \omega _n^2\right) \nonumber \\
  &+&
  16 \sqrt{14} y^3 z \left(761 \mu ^2-129 \omega _n^2\right)
  +
  14 y^2 z^2 \left(2171 \mu ^2+1701 \omega _n^2\right)+1344 \sqrt{14} y z^3 \left(\mu ^2+\omega _n^2\right)+49 z^4 \left(69 \mu ^2-\omega _n^2\right)\bigg)  \nonumber \\
  &+&
  40 \mu  p^2 \left(1172 y^4+338 \sqrt{14} y^3 z+196 y^2 z^2+105 \sqrt{14} y z^3-49 z^4\right) \left(\mu ^2+\omega _n^2\right) \\
  &+&
  35 \left(628 y^4+32 \sqrt{14} y^3 z-56 y^2 z^2+49 z^4\right) \left(\mu ^2+\omega _n^2\right){}^2+p^8 \big(-5996 y^4+384 \sqrt{14} y^3 z+27636 y^2 z^2
  \nonumber \\
  &+&5376 \sqrt{14} y z^3 - 5341 z^4\big)+8 \mu  p^6 \left(-1188 y^4+1542 \sqrt{14} y^3 z+9828 y^2 z^2+483 \sqrt{14} y z^3+637 z^4\right)
  \big] \nonumber
  \:,
 \end{eqnarray}
\begin{eqnarray}
 q_3&=& T \sum_n \int_0^\Lambda dp
 \frac{p^2}{154350 \pi ^2 \left(2 \omega _n^2 \left(\mu ^2+p^4\right)+\omega _n^4+\left(p^4-\mu ^2\right)^2\right){}^2}
 \big[
 y^4 \bigg(13321 \mu ^4+2 \omega _n^2 \left(13321 \mu ^2+20013 p^4+8256 \mu  p^2\right) \nonumber \\
 &+&
 13321 \omega _n^4-31823 p^8-77376 \mu  p^6-41714 \mu ^2 p^4+16512 \mu ^3 p^2\bigg)-48 \sqrt{14} y^3 z \big(133 \mu ^4+2 \omega _n^2 \left(133 \mu ^2+654 p^4+363 \mu  p^2\right) \nonumber \\
 &+&
 133 \omega _n^4+71 p^8+452 \mu  p^6+1028 \mu ^2 p^4+726 \mu ^3 p^2\big)-14 y^2 z^2 \bigg(9149 \mu ^4+2 \omega _n^2 \left(9149 \mu ^2-1023 p^4+7964 \mu  p^2\right) \\
 &+&
 9149 \omega _n^4+1813 p^8+4856 \mu  p^6+5934 \mu ^2 p^4+15928 \mu ^3 p^2\bigg)-168 \sqrt{14} y z^3 \bigg(7 \mu ^4+2 \omega _n^2 \left(7 \mu ^2+56 p^4+27 \mu  p^2\right) \nonumber \\
 &+&
 7 \omega _n^4+9 p^8+58 \mu  p^6+312 \mu ^2 p^4+54 \mu ^3 p^2\bigg)-98 z^4 \bigg(-217 \mu ^4-2 \omega _n^2 \left(217 \mu ^2+181 p^4-88 \mu  p^2\right)-217 \omega _n^4+71 p^8 \nonumber \\
 &+&
 752 \mu  p^6 - 222 \mu ^2 p^4+176 \mu ^3 p^2\bigg)
 \big] \nonumber
 \:.
\end{eqnarray}
\end{widetext}

At low temperatures we then find $q_1 >0$ and,
\begin{equation}
q_i = \frac{0.106 f_i (y)}{1543500 \pi^2} \frac{\sqrt{\mu}}{T_c ^2}
\end{equation}
with the functions $f_i (y)$ of the mixing as
\begin{widetext}
\begin{equation}
f_1(y)=34401 y^4+26436 \sqrt{14} y^3 z+111804 y^2 z^2+15624 \sqrt{14} y z^3+11956 z^4
 \:,
 \label{eq:q1numericalconstant}
\end{equation}
\begin{equation}
f_2(y)= 8 \big( 584 y^4+404 \sqrt{14} y^3 z+1351 y^2 z^2+126 \sqrt{14} y z^3+49 z^4 \big)
 \:,
 \label{eq:q2numericalconstant}
\end{equation}
\begin{equation}
 f_3(y) = -\frac{25}{2} \big(3027 y^4+2892 \sqrt{14} y^3 z+13188 y^2 z^2+1848 \sqrt{14} y z^3+1372 z^4 \big)
\:.
\label{eq:q3numericalconstant}
\end{equation}

% %
% %
% % At low temperatures we then find $q_1 >0$ and,
% % \begin{equation}
% % q_i = \frac{0.106 f_i (y)}{3200 \pi^2} \frac{\sqrt{\mu}}{T_c ^2}
% % \end{equation}
% % with the two functions of the mixing as
% %
% % \begin{equation}
% % f_2(y) = (74752 y^4 + 51712 \sqrt{14} y^3 z + 112168 y^2 z^2 + 16108\sqrt{14} y z^3 + 6272 z^4)/123480,
% % \end{equation}
% % \begin{equation}
% % f_3 (y) = -( 121080 y^4 + 115680 \sqrt{14} y^3 z + 527520 y^2 z^2 + 73920 \sqrt{14} y z^3 + 54880 z^4)/ 123480
% % \end{equation}
\end{widetext}
Evidently, $q_2 >0$ and $q_3 <0$, and the real states minimize the free energy for any mixing. This is also illustrated in Fig.~\ref{fig:q2q3coefficients}, where we vary the interactions in the range $V_1<0$ and $V_3 < 7 V_1 /27$ and plot the representative points, only to find them always in the lower right quadrant.

In search for the time-reversal-preserving  $j=2$ ground state we may first note that all real states can be rotated into a superposition of the biaxial and uniaxial states as
\begin{equation}
\frac{\Delta_1}{\sqrt{2}} \big( \left|2,+2 \right> + \left| 2, -2 \right> \big) + \Delta_2 \left|2,0\right>
\:,
\end{equation}
where $\Delta_1$ and $\Delta_2$ are both real, and $\Delta_1 ^2 + \Delta_2 ^2 =1$ \cite{boettcher2}. To determine the real superconducting ground state we study the sixth order terms in the GL free energy. Restricting ourselves to the real states we find it to be a sum of two terms:
\begin{equation}
 F^{j=2,real}_6
 =
 v_1 |\left< \Delta|\Delta\right>|^3 + v_2 \frac{4}{3} \Delta_2 ^2 (\Delta_2^2-3 \Delta_1^2)^2
 \:.
\end{equation}
The sign of the coefficient $v_2$ decides on the real superconducting ground state. If $v_2>0$, the biaxial nematic state with $\Delta_1=1$ and $\Delta_2=0$ has the lowest free energy. If $v_2<0$, on the other hand, the uniaxial nematic state with $\Delta_1=0$ and $\Delta_2=1$ minimizes the free energy. These two states also differ in their quasiparticle energy spectrum: in the case of the uniaxial nematic state we find point nodes along the $z$-axis, while the biaxial nematic state is fully gapped.

To establish the sign of $v_{2}$ one needs  to calculate the sixth-order term
\begin{equation}
 F_6(\Delta) = -\frac{32}{3} K_{abcdef} \Delta_a^* \Delta_b \Delta_c^* \Delta_d \Delta_e^* \Delta_f,
\end{equation}
with the one-loop integral as
 \begin{equation}
 \begin{split}
  K_{abcdef}= \mathrm{Tr} \int_Q^{\Lambda}
  &G_0(-\omega,-\textbf{p}) J_{2,m_a}^{ \dagger} G_0(\omega,\textbf{p}) J_{2,m_b} \\
  & \times
  G_0(-\omega,-\textbf{p}) J_{2,m_c}^{\dagger} G_0(\omega,\textbf{p}) J_{2,m_d} \\
  & \times
  G_0(-\omega,-\textbf{p}) J_{2,m_e}^{\dagger} G_0(\omega,\textbf{p}) J_{2,m_f}
  \:.
  \end{split}
 \end{equation}
At low temperature we find (see Appendix \ref{ap:leading_order_coeff})
\begin{equation}
v_2 = \frac{0.00773 f_v (y) \sqrt{\mu} }{ 2 \pi^2 T_c ^4},
\end{equation}
with
\begin{widetext}
\begin{equation}
 f_v (y) =
 -\frac{4595157 y^6+3369438 \sqrt{14} y^5 z+12731670 y^4 z^2+1076040 \sqrt{14} y^3 z^3-2081520 y^2 z^4-508032 \sqrt{14} y z^5-406112 z^6}{ 5793913125 }
 \:.
\end{equation}
\end{widetext}
Unlike the functions $f_i (y)$ in Eqs.~\eqref{eq:q1numericalconstant}--\eqref{eq:q3numericalconstant}, the function $f_v(y)$ changes sign. The change of sign occurs at $y=0.4547$, which corresponds to the ratio of the interaction parameters $|V_3|/|V_1| = 0.38445>7/27$, at which therefore there  is a further transition between the uniaxial and biaxial nematic states (see Fig.~\ref{fig:phase12}).

\subsection{$j=3$}
In this section we assume $\mu<0$, and determine the superconducting ground state when $j=3$ by using the previously derived general phase diagram for $j=3$ GL free energy of Ref.~\cite{kawaguchi}.

The fourth-order terms in the GL  free energy for $j=3$ are given by Eq.~\eqref{eq:f43}. The coefficients $c_i$ are given by the following one-loop expressions (Appendix \ref{ap:matching_conditions}):
\begin{widetext}
\begin{equation}
 c_1=T \sum_n \int_0^\Lambda dp
 \frac{4 p^2 \left(194 p^8-1677 p^6 \mu +715 p^4 \left(5 \mu ^2-\omega ^2\right)-3003 p^2 \mu  \left(\mu ^2+\omega ^2\right)+1001 \left(\mu ^2+\omega ^2\right)^2\right)}{75075 \pi ^2 \left(p^8-2 p^4 \left(\mu ^2-\omega ^2\right)+\left(\mu ^2+\omega ^2\right)^2\right)^2}
 \:,
\end{equation}
\begin{equation}
 c_2=T \sum_n \int_0^\Lambda dp
 \frac{p^2 \left(10337 p^8-45006 p^6 \mu +11440 p^4 \left(7 \mu ^2+3 \omega ^2\right)-68354 p^2 \mu  \left(\mu ^2+\omega ^2\right)+23023 \left(\mu ^2+\omega ^2\right)^2\right)}{900900 \pi ^2 \left(p^8-2 p^4 \left(\mu ^2-\omega ^2\right)+\left(\mu ^2+\omega ^2\right)^2\right)^2}
\:,
 \end{equation}
\begin{eqnarray}
\frac{ c_3}{7} =T \sum_n \int_0^\Lambda dp\frac{8 p^2 \left(479 p^8-2067 p^6 \mu +715 p^4 \left(5 \mu ^2+3 \omega ^2\right)-3003 p^2 \mu  \left(\mu ^2+\omega ^2\right)+1001 \left(\mu ^2+\omega ^2\right)^2\right)}{225225 \pi ^2 \left(p^8-2 p^4 \left(\mu ^2-\omega ^2\right)+\left(\mu ^2+\omega ^2\right)^2\right)^2}
\:,
 \end{eqnarray}
\begin{eqnarray}
  \frac{c_4}{7} =T \sum_n \int_0^\Lambda dp \frac{p^2 \left(9977 p^8-38454 p^6 \mu +13156 p^4 \left(5 \mu ^2+3 \omega ^2\right)-55770 p^2 \mu  \left(\mu ^2+\omega ^2\right)+19019 \left(\mu ^2+\omega ^2\right)^2\right)}{750750 \pi ^2 \left(p^8-2 p^4 \left(\mu ^2-\omega ^2\right)+\left(\mu ^2+\omega ^2\right)^2\right)^2}
  \:.
\end{eqnarray}
\end{widetext}
As before, after performing the sum over Matsubara frequencies and the momentum integral, at low temperatures we find
\begin{equation}
 c_1= \frac{72}{143} \frac{0.10657}{32 \pi^2} \frac{\sqrt{|\mu|} } { T_c^2 }= 2 c_2 = \frac{c_3}{5} = \frac{2 c_4}{7}
 \:.
\end{equation}
We find therefore that $c_1 >0$, and $c_3/c_1 = 5$, $c_4 / c_1 = 7/2$, and $c_3/c_4 = 10/7$.
As can be seen in Fig.~6 (a) of Ref. \cite{kawaguchi}, for example, these numbers place the system a bit below the phase boundary between the phases ``E" and ``D", which lies at $c_3/c_4 = 5/3 $ in our notation. The $j=3$ ground state is therefore the phase ``D", i. e.
\begin{equation}
\frac{1}{\sqrt{2}} \big( \left| 3,+2 \right> - \left| 3, -2\right> \big).
\end{equation}
This superconducting condensate is symmetric under cubic transformations, and respects time reversal. The quasiparticle energy spectrum of this state exhibits   six gapless points at $ (\mu^2 + (\Delta^2 / 12) )^{\frac{1}{4}} \{ (0,0,\pm 1), (0,\pm 1, 0), (\pm 1,0,0) \}$.

\section{Summary and Discussion}
\label{sec:sumdis}

In conclusion, we obtained the phase diagram  of the rotationally invariant Luttinger semimetal with weak attraction in the $l=1$ ($p$-wave) channel. The total angular momentum of the superconducting phases that appear at low temperatures depends on the sign of the chemical potential, with the further selection of the ground state provided by the fourth-order and the sixth-order terms in the GL free energy. While the residual spatial symmetry of the five possible
ground-state condensates varies, the feature common to all is the preservation of the time-reversal symmetry, and the absence of nodal lines in the quasiparticle spectrum.

When the pairing interaction is spin independent, the two interaction parameters are equal; $V_1 = V_3$, and the superconducting ground state is $|0,0\rangle$ on the hole-doped, and $|2,0\rangle$ on the particle-doped side. In either case the quasiparticle spectrum features the full gap, albeit an  anisotropic one in the latter case. This may be contrasted with the condensate with $s=2$ that results from the attraction in the $l=0$ channel \cite{boettcher2}, which at $\mu=0$ at least has the same quantum numbers $j=2$ and $m_j=0$, and the same symmetry, but as all other real states in that case exhibits lines of gapless excitations. Furthermore, when the weak attraction is in the $l=0$ (or $l=2$) channel, the phase diagram is independent of the sign of chemical potential, and the $s=2$ ($s=0$) ground state at a finite chemical potential breaks time-reversal symmetry \cite{boettcher2, herbut2}. All of these features stand in stark contrast to the $p$-wave ground states we discussed in this paper. On the other hand, the preservation of the time reversal and the concomitant full gap in the excitation spectrum was found before in the $SO_L (3) \times SO_S (3)$ - symmetric GL free energy for the $l=1$, $s=1$ matrix-order parameter that pertains to $^3 He$, when the order parameter is restricted to $j=2$ \cite{sauls}. This result resembles what we find quite generally for the $p$-wave states in the Luttinger semimetal.

While we find the quasiparticle spectra to be either gapless or with a full gap, depending on the particular $p$-wave ground state, no ground state showed a line of gapless points. Our explicit computation of the energy spectra is in agreement with general arguments of the ref. \cite{venderbos}.  At weak coupling there is in general always just one particular value of $j$ that becomes favored below the critical temperature, i. e. the condensate is never a linear combination of states with different $j$, unless the system is accidentally right  at the boundary between two different phases. We have not checked the quasiparticle spectrum at such special cases of attractive interaction. It is not inconceivable that some such linear combinations may yield lines of gapless points, as suggested by the penetration depth data in YPtBi \cite{kim} for example, but this would seem to require special tuning. Generic minima of the weak-coupling GL free energy, both in the cases $(l=0,s=2,j=2)$ and $(l=2,s=0,j=2)$ ($d$-wave) studied earlier, and in the present case of general attraction in the $l=1$ case ($p$-wave) do not show this feature, however.

We expect our results for the competition between states with different $j$ to be typically in agreement with the weak-coupling RG flows, as it is the case with many other weak-coupling mean-field treatments of competing instabilities. There could be exemptions, however, such as in the case of mixing of two $j=2$ channels, for example.

\section{Acknowledgement}

J.M.L. is supported by the DFG grant No. LI 3628/1-1, and and I.F.H. by the NSERC of Canada.

\appendix

\section{The Luttinger Hamiltonian and the Green's function}
\label{app:LuttAndGreen}

The celebrated Luttinger-Kohn Hamiltonian  is \cite{luttinger}:
\begin{equation}
 H_0= \frac{1}{2m} \left(\frac{5}{4} p^2 - (\boldsymbol{p}\cdot \boldsymbol{\Sigma})^2 \right) + \frac{p^2}{2 m_0}-\frac{p_a^2 \Sigma_a^2}{2 m_1} -\mu
 \:,
\end{equation}
where the $\boldsymbol{\Sigma}=(\Sigma_x,\Sigma_y,\Sigma_z)^{\rm T}$-matrices are the spin--$3/2$ representation of the Lie algebra of $SO(3)$ and have the form
\begin{equation}
 \Sigma_x = \begin{pmatrix}
      0 & \sqrt{3}/2 & 0 & 0\\
      \sqrt{3}/2 & 0 & 1 & 0\\
      0 & 1 & 0 & \sqrt{3}/2\\
      0 & 0 & \sqrt{3}/2 & 0
     \end{pmatrix}\:,
\end{equation}
\begin{equation}
 \Sigma_y = \frac{i}{2}\begin{pmatrix}
        0 & -\sqrt{3} & 0 & 0 \\
        \sqrt{3} & 0 & -2 & 0\\
        0 & 2 & 0 & -\sqrt{3} \\
        0 & 0 & \sqrt{3} & 0
       \end{pmatrix}
\:,
\end{equation}
\begin{equation}
\Sigma_z = \frac{1}{2} \begin{pmatrix}
                    3 & 0 & 0& 0\\
                    0 & 1 & 0 & 0\\
                    0 & 0 & -1 & 0 \\
                    0 & 0 & 0 & -3
                   \end{pmatrix}
\:.
\end{equation}
This Hamiltonian can be rewritten in terms of the real $\ell=2$ spherical harmonics $d_a(\textbf{p})$, and the Dirac matrices $\gamma_a$, which obey the Clifford algebra $\{\gamma_a, \gamma_b \}=2 \delta_{ab}$ \cite{abrikosov}:
\begin{equation}
 H_0= \sum_{a=1}^5\frac{d_a(\textbf{p}) \gamma_a}{2 m} + \frac{p^2}{2 m_0} +\frac{d_1(\textbf{p}) \gamma_1 + d_2(\textbf{p}) \gamma_2}{2 m_1} -\mu
 \:.
\end{equation}
The spherical harmonics are given by
\begin{eqnarray*}
 d_1(\textbf{p})&=& \frac{\sqrt{3}}{2} (p_x^2-p_y^2)\:,\quad d_2(\textbf{p})= \frac{1}{2} (2 p_z^2-p_x^2-p_y^2)\:, \\
 d_3(\textbf{p})&=&\sqrt{3} p_x p_z \:, \quad
 d_4(\textbf{p})= \sqrt{3} p_y p_z\:, \quad
 d_5(\textbf{p}) = \sqrt{3} p_x p_y\:,
\end{eqnarray*}
while the corresponding Dirac matrices are:
\begin{widetext}
 \begin{eqnarray}
  \gamma_1 &=& \begin{pmatrix}
              0 & 0 & 1 & 0\\
              0 & 0 & 0 & 1\\
              1 & 0 & 0 & 0\\
              0 & 1 & 0 & 0
             \end{pmatrix}\:, \quad
    \gamma_2 =\begin{pmatrix}
               1 & 0 & 0 & 0 \\
               0 & -1 & 0 & 0\\
               0 & 0 & -1 & 0\\
               0 & 0 & 0 & 1
              \end{pmatrix}\:, \\
    \gamma_3 &=&\begin{pmatrix}
                 0 & 1 & 0 & 0\\
                 1 & 0 & 0 & 0\\
                 0 & 0 & 0 & -1\\
                 0 & 0 & -1 & 0\\
                \end{pmatrix}\:, \quad
    \gamma_4 = \begin{pmatrix}
                0 & -i & 0 & 0 \\
                i & 0 & 0 & 0\\
                0 & 0 & 0 & i \\
                0 & 0 & -i & 0
               \end{pmatrix}\:, \quad
    \gamma_5 = \begin{pmatrix}
                0 & 0 & -i & 0 \\
                0 & 0 & 0 & -i\\
                i & 0 & 0 & 0\\
                0 & i & 0 & 0
               \end{pmatrix}
\:.
 \end{eqnarray}
\end{widetext}
One advantage of expressing the Hamiltonian in terms of the  Dirac matrices and spherical harmonics is that the Green's function has a simple analytic expression, which is
\begin{equation}
 G_0(\omega,\textbf{p})=
 \left( i\omega -H_0 \right)^{-1}
 =\frac{i \omega - d_a(\textbf{p}) \gamma_a + \mu}{(i\omega + \mu)^2 -p^4}
 \:.
\end{equation}
where we set $m_0 = m_1 = \infty$, and $2 m=1$.

Kohn-Luttinger Hamiltonian commutes with the antiunitary time-reversal operator $T=U \mathcal{K}$, which consists of the unitary matrix $U$ and complex conjugation $\mathcal{K}$. The unitary part of the time-reversal operator in the above representation is defined as \cite{boettcher1}
\begin{equation}
 U=i \gamma_4 \gamma_5
 \:.
\end{equation}
Evidently, $T ^2 = -1$, and the Kohn-Luttinger Hamiltonian describes a fermion with half-integer spin.

\section{Pairing matrices}
\label{app:interations}
In this section we provide explicit expressions of the pairing matrices $J^{(k)}_{j,m_j}$.
The pairing matrices are defined as
 \begin{equation}
  J^{(k)}_{j,m_j}(\textbf{p})=
  \sum_{m_l +m_k=m_j} \left< L_1 m_l,S_k m_k | j m_j \right> L_{1,m_l}(\textbf{p}) S_{k,m_k}
  \:,
 \end{equation}
where $\left< L_1 m_l,S_k m_k | j m_j \right>$ are the Clebsch-Gordan coefficients and $L_{1,m_l}$ are the spherical harmonics of $L=1$. The spherical harmonics are given by
\begin{eqnarray}
 L_{1,+1} &=&-\frac{1}{\sqrt{2}}\frac{p_x + i p_y}{p} \:, \\
 L_{1,0} &=& \frac{p_z}{p} \:, \\
 L_{1,-1} &=& \frac{1}{\sqrt{2}} \frac{p_x -i p_y}{p} \:.
\end{eqnarray}
The matrices $S_{k,m_k}$ denoting the spin $s=1$ or $s=3$ of the Cooper pair are defined in the following subsection.
\subsection{$s=1$ and $s=3$ matrices }
To find the matrices for $s=1$ and $s=3$ spin, we define first the ladder operators \cite{yang}:
\begin{eqnarray}
 S_{+}& =&(\Sigma_x + i \Sigma_y) \:, \\
 S_- &=& (\Sigma_x - i \Sigma_y) \:.
\end{eqnarray}
With the help of the ladder operators, the three matrices for $s=1$ can be defined as:
\begin{eqnarray}
 S_{1,1}&=& -\frac{1}{\sqrt{2}} S_+ \\
 S_{1,0}&=& \Sigma_z \\
 S_{1,-1}&=& \frac{1}{\sqrt{2}} S_- \:.
\end{eqnarray}
The seven matrices for $s=3$ can be then obtained by using the fact that they transform as a third-rank irreducible tensor under $SO(3)$:
\begin{equation}
 [S_-, S_{j,m}]= \sqrt{(j+m)(j-m+1)} S_{j,m-1}
 \:,
\end{equation}
which yields
\begin{eqnarray}
 S_{3,+3} &=& \frac{\sqrt{2}}{3} (S_{1,1})^3 \:, \\
 S_{3,+2} &=& \frac{1}{\sqrt{6}} [S_-,S_{3,+3}] \:, \\
 S_{3,+1} &=& \frac{1}{\sqrt{10}} [S_-, S_{3,+2}] \:, \\
 S_{3,0} &=& \frac{1}{2 \sqrt{3}} [S_-, S_{3,+1}] \:, \\
 S_{3,-1} &=& -S_{3,+1}^\dagger \:, \\
 S_{3,-2} &=& -S_{3,+2}^\dagger  \:, \\
 S_{3,-3} &=& -S_{3,+3}^\dagger
 \:.
\end{eqnarray}

We then find the following expressions for the pairing matrices:

\subsection{(1,1,j)-channel}

$j=0$ :

\begin{equation}
\textstyle
 J^{(\mathit{1})}_{0,0}
 =\frac{1}{\sqrt{15}}(L_{1,-1} S_{1,+1}+ L_{1,+1} S_{1,-1} -L_{1,0} S_{1,0} ).
\end{equation}

$j=1$ :

\begin{eqnarray}
 J^{(\mathit{1})}_{1,1} &=&
 \frac{1}{\sqrt{2}} \frac{1}{\sqrt{5}}(L_{1,+1} S_{1,0}- L_{1,0} S_{1,+1})\\
 %&=&
 %\frac{1}{\sqrt{2}}\frac{1}{\sqrt{5}} (\epsilon_3 - i \epsilon_1),  \\
 J^{(\mathit{1})}_{1,0} &=&
 \frac{1}{\sqrt{2}}\frac{1}{\sqrt{5}} (L_{1,+1} S_{1,-1} - L_{1,-1} S_{1,+1})\\
% &=& i\frac{1}{\sqrt{5}} \epsilon_2, \\
 J^{(\mathit{1})}_{1,-1}
 &=& \frac{1}{\sqrt{2}}\frac{1}{\sqrt{5}} (L_{1,0} S_{1,-1} - L_{1,-1} S_{1,0})%\\
% &=& \frac{1}{\sqrt{2}} \frac{1}{\sqrt{5}}(\epsilon_3 + i \epsilon_1).
\end{eqnarray}

$j=2$:

\begin{eqnarray}
\textstyle
 J^{(\mathit{1})}_{2,+2}
 \textstyle &=& \textstyle \frac{1}{\sqrt{5}}
 L_{1,1} S_{1,1},\\
 \textstyle J^{(\mathit{1})}_{2,+1} &=& \textstyle \frac{1}{\sqrt{2}}\frac{1}{\sqrt{5}} (L_{1,0} S_{1,1} + L_{1,+1} S_{1,0}),\\
 \textstyle J^{(\mathit{1})}_{2,0} &=& \scriptstyle
 \frac{1}{\sqrt{30}} (L_{1,1} S_{1,-1} +2 L_{1,0} S_{1,0} +L_{1,-1} S_{1,1}), \\
 \textstyle J^{(\mathit{1})}_{2,-1} &=& \textstyle \frac{1}{\sqrt{2}}\frac{1}{\sqrt{5}} (L_{1,0} S_{1,-1} + L_{1,-1} S_{1,0}),\\
\textstyle J^{(\mathit{1})}_{2,-2}
 &=& \textstyle
 \frac{1}{\sqrt{5}} L_{1,-1} S_{1,-1}.
\end{eqnarray}
\subsection{$(1,3,j)$-channel}
$j=2$:
\begin{eqnarray}
 \scalemath{0.73}{J^{(\mathit{3})}_{2,+2}}
  &\scalemath{0.73}{=}&
 \scalemath{0.73}{\sqrt{\frac{5}{7}} L_{1,-1} S_{3,+3} -\sqrt{\frac{5}{21}} L_{1,0} S_{3,+2} + \sqrt{ \frac{1}{21}} L_{1,+1} S_{3,+1}}\\
 \scalemath{0.73}{ J^{(\mathit{3})}_{2,+1} }
  &\scalemath{0.73}{=}&
 \scalemath{0.73}{ \sqrt{\frac{10}{21}} L_{1,-1} S_{3,+2} -2 \sqrt{\frac{2}{21}} L_{1,0} S_{3,+1} +\frac{1}{\sqrt{7}} L_{1,+1} S_{3,0} }\\
  \scalemath{0.73}{J^{(\mathit{3})}_{2,0}}
   &\scalemath{0.73}{=}&
  \scalemath{0.73}{\frac{1}{\sqrt{7}} (\sqrt{2} L_{1,-1} S_{3,+1} -\sqrt{3} L_{1,0} S_{3,0} + \sqrt{2} L_{1,+1} S_{3,-1}) } \\
  \scalemath{0.73}{J^{(\mathit{3})}_{2,-1}}
  &\scalemath{0.73}{=}&
 \scalemath{0.73}{\sqrt{\frac{10}{21}} L_{1,+1} S_{3,-2} -2 \sqrt{\frac{2}{21}} L_{1,0} S_{3,-1} +\frac{1}{\sqrt{7}} L_{1,-1} S_{3,0} }\\
 \scalemath{0.73}{J^{(\mathit{3})}_{2,-2} }
  &\scalemath{0.73}{=}&
 \scalemath{0.73}{\sqrt{\frac{5}{7}} L_{1,+1} S_{3,-3} -\sqrt{\frac{5}{21}} L_{1,0} S_{3,-2} + \sqrt{ \frac{1}{21}} L_{1,-1} S_{3,-1},}
\end{eqnarray}
$j=3$ :

\begin{eqnarray}
\scalemath{0.75}{J^{(\mathit{3})}_{3,+3}}
  &\scalemath{0.75}{=}&
 \scalemath{0.75}{ \frac{1}{2} (L_{1,+1} S_{3,+2} -\sqrt{3} L_{1,0} S_{3,+3})}\\
  \scalemath{0.75}{J^{(\mathit{3})}_{3,+2}}
  &\scalemath{0.75}{=}&
  \scalemath{0.75}{\frac{1}{2}\sqrt{\frac{5}{3}} L_{1,+1} S_{3,+1}-\frac{1}{\sqrt{3}} L_{1,0} S_{3,+2} -\frac{1}{2} L_{1,-1} S_{3,3}}\\
  \scalemath{0.75}{J^{(\mathit{3})}_{3,+1} }
  &\scalemath{0.75}{=}&
  \scalemath{0.75}{-\frac{\sqrt{\frac{5}{3}}}{2} L_{1,-1} S_{3,+2} -\frac{1}{2 \sqrt{3}}  L_{1,0} S_{3,+1}+ \frac{1}{\sqrt{2}}  L_{1,+1} S_{3,0} }\\
  \scalemath{0.75}{J^{(\mathit{3})}_{3,0}}
  &\scalemath{0.75}{=}&
  \scalemath{0.75}{\frac{1}{\sqrt{2}} (L_{1,+1} S_{3,-1} - L_{1,-1} S_{3,+1}) }\\
  \scalemath{0.75}{ J^{(\mathit{3})}_{3,-1} }
  &\scalemath{0.75}{=}&
 \scalemath{0.75}{ \frac{\sqrt{\frac{5}{3}}}{2} L_{1,+1} S_{3,-2}+\frac{1}{2 \sqrt{3}} L_{1,0} S_{3,-1} -\frac{1}{\sqrt{2}} L_{1,-1} S_{3,0} }\\
   \scalemath{0.75}{  J^{(\mathit{3})}_{3,-2}}
 &\scalemath{0.75}{=}&
 \scalemath{0.75}{ \frac{1}{2} L_{1,+1} S_{3,-3}+ \frac{1}{\sqrt{3}} L_{1,0} S_{3,-2} -\frac{\sqrt{\frac{5}{3}}}{2} L_{1,-1} S_{3,-1} }\\
    \scalemath{0.75}{   J^{(\mathit{3})}_{3,-3}}
  &\scalemath{0.75}{=}&
 \scalemath{0.75}{ \frac{\sqrt{3}}{2} L_{1,0} S_{3,-3} - \frac{1}{2} L_{1,-1} S_{3,-2},}
\end{eqnarray}
$j=4$

\begin{eqnarray}
   \scalemath{0.73}{J^{(\mathit{3})}_{4,+4}}
   &\scalemath{0.73}{=}&
   \scalemath{0.73}{L_{1,+1} S_{3,+3} } \\
   \scalemath{0.73}{ J^{(\mathit{3})}_{4,+3} }
   &\scalemath{0.73}{=}&
   \scalemath{0.73}{\frac{1}{2}(L_{1,0}S_{3,+3}+ \sqrt{3} L_{1,+1} S_{3,+2}) }\\
   \scalemath{0.73}{ J^{(\mathit{3})}_{4,+2}}
   &\scalemath{0.73}{=}&
   \scalemath{0.73}{\frac{1}{\sqrt{7}} (\frac{1}{2} L_{1,-1} S_{3,+3}+ \sqrt{3} L_{1,0} S_{3,+2} +\frac{\sqrt{15}}{2} L_{1,+1} S_{3,+1}) }\\
   \scalemath{0.73}{ J^{(\mathit{3})}_{4,+1} }
   &\scalemath{0.73}{=}&
   \scalemath{0.73}{ \frac{\sqrt{\frac{3}{7}}}{2} L_{1,-1} S_{3,+2}+\frac{\sqrt{\frac{15}{7}}}{2} L_{1,0} S_{3,+1}+\sqrt{\frac{5}{14}} L_{1,+1} S_{3,0} } \\
   \scalemath{0.73}{ J^{(\mathit{3})}_{4,0} }
   &\scalemath{0.73}{=}&
   \scalemath{0.73}{\sqrt{\frac{3}{14}} L_{1,-1} S_{3,+1} + \frac{2}{\sqrt{7}} L_{1,0} S_{3,0}+ \sqrt{\frac{3}{14}} L_{1,+1} S_{3,-1} }\\
   \scalemath{0.73}{ J^{(\mathit{3})}_{4,-1} }
   &\scalemath{0.73}{=}&
    \scalemath{0.73}{\frac{\sqrt{\frac{3}{7}}}{2} L_{1,+1} S_{3,-2}+\frac{\sqrt{\frac{15}{7}}}{2} L_{1,0} S_{3,-1}+\sqrt{\frac{5}{14}} L_{1,-1} S_{3,0} }\\
    \scalemath{0.73}{J^{(\mathit{3})}_{4,-2}}
    &\scalemath{0.73}{=}&
    \scalemath{0.73}{\frac{1}{\sqrt{7}} (\frac{1}{2} L_{1,+1} S_{3,-3}+ \sqrt{3} L_{1,0} S_{3,-2} +\frac{\sqrt{15}}{2} L_{1,-1} S_{3,-1}) }\\
   \scalemath{0.73}{ J^{(\mathit{3})}_{4,-3} }
   &\scalemath{0.73}{=}&
   \scalemath{0.73}{\frac{1}{2}(L_{1,0}S_{3,-3}+ \sqrt{3} L_{1,-1} S_{3,-2})}\\
   \scalemath{0.73}{ J^{(\mathit{3})}_{4,-4} }
    &\scalemath{0.73}{=}&
    \scalemath{0.73}{L_{1,-1} S_{3,-3}}
 \end{eqnarray}

% % % and in  real basis,
% % %  \begin{eqnarray}
% % %   \alpha_1 &=& J^{(\mathit{3})}_{4,0}, \\
% % %   \alpha_2 &=& \frac{1}{\sqrt{2}} (J^{(\mathit{3})}_{4,-4} + J^{(\mathit{3})}_{4,+4}),\\
% % %   \alpha_3 &=& \frac{i}{\sqrt{2}} (J^{(\mathit{3})}_{4,-4} - J^{(\mathit{3})}_{4,+4}),\\
% % %   \alpha_4 &=& \frac{i}{\sqrt{2}} (J^{(\mathit{3})}_{4,-3} + J^{(\mathit{3})}_{4,+3}),\\
% % %   \alpha_5 &=& \frac{1}{\sqrt{2}} (J^{(\mathit{3})}_{4,-3} - J^{(\mathit{3})}_{4,+3}),\\
% % %   \alpha_6 &=& \frac{1}{\sqrt{2}} (J^{(\mathit{3})}_{4,-2} + J^{(\mathit{3})}_{4,+2}),\\
% % %   \alpha_7 &=& \frac{i}{\sqrt{2}} (J^{(\mathit{3})}_{4,-2} - J^{(\mathit{3})}_{4,+2}),\\
% % %   \alpha_8 &=& \frac{i}{\sqrt{2}} (J^{(\mathit{3})}_{4,-1} + J^{(\mathit{3})}_{4,+1}),\\
% % %   \alpha_9 &=& \frac{1}{\sqrt{2}} (J^{(\mathit{3})}_{4,-1} - J^{(\mathit{3})}_{4,+1}).
% % %  \end{eqnarray}

\section{Leading-order calculation of the coefficients}
\label{ap:leading_order_coeff}

\subsection{Second-order coefficient}
The one-loop integrals $K_{j,ab}^{(s_1 s_2)}$ defined in Eq.~\eqref{eq:secondOrderKab} have the following structure:
\begin{equation}
 K_{j,ab}= T \sum_n \int_0^\Lambda p^2 dp \frac{f_{j,ab}(p,\omega_n)}{[(p^2-\mu)^2+\omega_n^2][(p^2+\mu)^2+\omega_n^2]}
\end{equation}
which can be approximated around $p=\sqrt{|\mu|}$ and $\omega_n=0$ as
\begin{equation}
 K_{j,ab}= \frac{ f_{j,ab}(\sqrt{|\mu|},0)}{4 \mu^2} T\sum_n \int_0^\Lambda p^2 dp \frac{1}{[(p^2-|\mu|)^2+\omega_n^2]}
 \:.
\end{equation}
After performing the Matsubara sum, we obtain
\begin{eqnarray}
 K_{j,ab}&=& \scalemath{0.7}{\frac{ f_{j,ab}(\sqrt{|\mu|},0)}{8 \mu^2} \int_0^\Lambda dp  \frac{p^2 \tanh\left| \frac{p^2-|\mu|}{2T} \right| }{ \left| p^2 - |\mu|  \right|} }\\
 &=& \scalemath{0.7}{\frac{ f_{j,ab}(\sqrt{|\mu|},0)}{8 \mu^2}\bigg[ \Lambda + \sqrt{|\mu|} \int_0^{\Lambda/|\mu|} dp \bigg( \frac{p^2 \tanh\left| \frac{|\mu|}{T}\frac{p^2-1}{2} \right| }{ \left| p^2 -1 \right|} -1 \bigg) \bigg] } \nonumber
\end{eqnarray}
and use
\begin{equation}
 \int_0^\infty \big[ \frac{z^2}{|z^2-1|} \tanh \frac{y |z^2-1|}{2} -1 \big]
 \to
 \log \big( \frac{8 e^{\gamma-2} }{\pi} y \big)
 \:
\end{equation}
for $y\to \infty$.
This leads to Eq.~\eqref{eq:secondOrderKabResult}
 \begin{eqnarray}
  K_{j, ab}^{(s_1 s_2 )}
 &=&\frac{c_{j, ab}^{(s_1 s_2 )}}{\pi^2} \bigg( \frac{1}{V_c} + \sqrt{|\mu|} \log\frac{|\mu|}{T} \bigg)
  \:,
  \label{eq:secondOrderKabResultb}
 \end{eqnarray}
with the non-universal critical interaction
\begin{equation}
 \frac{1}{V_c}=\Lambda+\sqrt{|\mu| } \log \frac{8 e^{\gamma -2}}{\pi}
\end{equation}
and the numerical coefficient
\begin{equation}
 c_{j,ab}^{(s_1 s_2)} = \frac{ f_{j,ab}(\sqrt{|\mu|},0) \pi^2}{8 \mu^2}
 \:.
\end{equation}

\subsection{Fourth-order coefficients}
All coefficients $a_i$ have the following structure:
\begin{eqnarray}
 a_i
 &=& T \sum_n \int_0^\Lambda dp \frac{f_i\big(p,\omega_n\big)}{\big(p^8-2 p^4(\mu^2-\omega_n^2)+(\mu^2+\omega_n^2)^2\big)^2}\\
 &\approx&
 T \sum_n \int_{-\epsilon}^\epsilon \frac{d\xi}{2 \sqrt{|\mu|}} \frac{f_i\big(p=\sqrt{|\mu|},\omega_n=0\big)}{16 \mu^4 (\xi^2 +\omega_n^2)^2}
 \:,
\end{eqnarray}
where the change of variable $\xi = p^2-|\mu|$ was made. After performing the Matsubara sum for finite temperatures, we find:
\begin{equation}
 a_i= \frac{f_i(\sqrt{|\mu|},0)}{32 |\mu| ^{9/2}} \frac{1}{T^2}\int_{-\epsilon/T}^{\epsilon/T} d\hat{\xi} \:g(\hat{\xi}) \:,
\end{equation}
with $\hat{\xi}=\xi/T$ and
\begin{equation}
 g(z) =\frac{\sinh z-z}{8 z^3 \cosh^2 (z/2)}
 \:.
\end{equation}
At low temperatures the integral converges  and one obtains:
\begin{equation}
 a_i= \frac{f_i(\sqrt{|\mu|},0)}{32 |\mu| ^{9/2}} \frac{0.10657}{T_c^2}
 \:.
\end{equation}
\subsection{Sixth-order coefficients}
The two coefficients $v_1$ and $v_2$ possess the following structure:
\begin{equation}
 v_i= T \sum_n \int_0^\Lambda \frac{\tilde{f}_i\big(p, \omega_n \big)}{\left(2 \omega _n^2 \left(\mu ^2+p^4\right)+\omega _n^4+\left(p^4-\mu ^2\right)^2\right)^3}
\end{equation}
In the weak-coupling limit, the leading-order result in small parameter $T_c/\mu$ is determined by
\begin{eqnarray}
 v_i & \approx & \tilde{f}_i(\sqrt{\mu},0) T \sum_n \int_{-\epsilon}^\epsilon \frac{d \xi}{2\sqrt{\mu}} \frac{1}{64 \mu^6 (\xi^2 +\omega_n^2)^3} \\
 &=& \frac{\tilde{f}_i(\sqrt{\mu},0)}{64 \mu^{13/2}} \frac{1}{2 T^5} \int_{-\epsilon/T}^{\epsilon/T} d\hat{\xi} \: \tilde{g}(\hat{\xi})
\end{eqnarray}
with $\hat{\xi}=\xi/T$ and
\begin{equation}
 \tilde{g}(z)= \frac{3 \sinh (z)-z \left(z \tanh \left(\frac{z}{2}\right)+3\right)}{16  z^5 (\cosh (z)+1)}\:.
\end{equation}
For small temperatures we therefore find
\begin{equation}
 v_i=\frac{\tilde{f}_i(\sqrt{\mu},0)}{64 \mu^{13/2}} \frac{0.00773}{2 T_c^5} \:.
\end{equation}
\section{Matching conditions}
\label{ap:matching_conditions}

 The quartic order of the Ginzburg-Landau free energy is defined by the one-loop integral
 \begin{equation}
 F_4^{\mathcal{J}}(\Delta)= 4  K_{abcd}^{(s_1 s_2 s_3 s_4)} \Delta_a^{(s_1)*} \Delta_b^{(s_2)} \Delta_c^{(s_3)*} \Delta_d^{(s_4)}
 \label{eq:quarticorderGLa}
 \end{equation}
with
\begin{widetext}
 \begin{equation}
 \label{eq:quarticorderloop}
  K_{abcd}^{(s_1 s_2 s_3 s_4)}= \mathrm{Tr} \int_Q^{\Lambda}
  G_0(-\omega,-\textbf{p}) J_{j,m_a}^{(s_1) \dagger}(\textbf{p}) G_0(\omega,\textbf{p}) J_{j,m_b}^{(s_2)}(\textbf{p}) \\
  G_0(-\omega,-\textbf{p}) J_{j,m_c}^{(s_3) \dagger} (\textbf{p}) G_0(\omega,\textbf{p}) J_{j,m_d}^{(s_4)}(\textbf{p})
  \:.
 \end{equation}
 \end{widetext}

 In the next section we demonstrate how the above expression is related to Eqs.~\eqref{eq:f41}--\eqref{eq:f43}.

 \subsection{$j=1$}
 To find the sign and magnitude of $\lambda_{1,2}$ defined in Eq.~\eqref{eq:f41}, we evaluate Eq.~\eqref{eq:quarticorderloop} for two different states. The first state is the real state $\left| 1,0 \right>$ with the pairing matrix $J^{(1)}_{1,0}$, while the second state is $\left|1,+1\right>$ with the pairing matrix $J^{(1)}_{1,+1}$. We find
 \begin{widetext}
 \begin{equation}
  F_4^1(\Delta_{\left| 1,0\right>})=
  T \sum_n \int_0^\Lambda dp \frac{p^2 \left(\omega _n^2 \left(82 \mu ^2-38 p^4+80 \mu  p^2\right)+41 \omega _n^4-\left(\mu +p^2\right)^2 \left(-41 \mu ^2+7 p^4+2 \mu  p^2\right)\right)}{375 \pi ^2 \left(2 \omega _n^2 \left(\mu ^2+p^4\right)+\omega _n^4+\left(p^4-\mu ^2\right)^2\right){}^2}
 \end{equation}
and
\begin{equation}
 F_4^1(\Delta_{\left|1,+1\right>})=
 T \sum_n \int_0^\Lambda dp \frac{p^2 \left(\omega _n^2 \left(42 \mu ^2-38 p^4+40 \mu  p^2\right)+21 \omega _n^4+\left(\mu +p^2\right)^2 \left(21 \mu ^2+13 p^4-2 \mu  p^2\right)\right)}{450 \pi ^2 \left(2 \omega _n^2 \left(\mu ^2+p^4\right)+\omega _n^4+\left(p^4-\mu ^2\right)^2\right){}^2}
 \:.
\end{equation}
\end{widetext}
Upon inserting these two states into Eq.~\eqref{eq:f41}, we obtain the following matching conditions:
\begin{eqnarray}
 F_4^1(\Delta_{\left| 1,0\right>})&=& \lambda_1 \\
 F_4^1(\Delta_{\left|1,+1\right>})&=& \lambda_1 +\lambda_2
 \:,
\end{eqnarray}
which yield the Eqs.~\eqref{eq:lambda1} and \eqref{eq:lambda2}.

\subsection{$j=2$}
To determine the coefficients $q_i$, we choose the states $\left|2,0\right>$, $\left|2,2\right>$ and $\left|2,1\right>$ and define the coefficient $q_\alpha=q_3/5$. The matching conditions of these states are given by
\begin{eqnarray}
 F_4^2(\Delta_{\left|2,0\right>}) &=& q_1 + q_\alpha ,\\
 F_4^2(\Delta_{\left|2,2\right>}) &=& q_1 + 4 q_2 ,\\
 F_4^2(\Delta_{\left|2,1\right>}) &=& q_1 + q_2 ,
\end{eqnarray}
which yields
\begin{eqnarray}
 q_1 &=& \scalemath{0.85}{\frac{1}{3} (-F_4^2(\Delta_{\left|2,2\right>})+4F_4^2(\Delta_{\left|2,1\right>})) } ,\\
 q_2 &=& \scalemath{0.85}{ \frac{1}{3} ( F_4^2(\Delta_{\left|2,2\right>})-F_4^2(\Delta_{\left|2,1\right>})) },\\
 q_\alpha &=& \scalemath{0.85}{ \frac{1}{3} [3 F_4^2(\Delta_{\left|2,0\right>})+F_4^2(\Delta_{\left|2,2\right>}) -4 F_4^2(\Delta_{\left|2,1\right>})] ,}
\end{eqnarray}
 with
 \begin{widetext}
  \begin{eqnarray}
   F_4^2(\Delta_{\left|2,0\right>}) &=& T \sum_n \int_0^\Lambda dp \frac{p^2}{771750 \pi ^2 \left(2 \omega _n^2 \left(\mu ^2+p^4\right)+\omega _n^4+\left(p^4-\mu ^2\right)^2\right){}^2}\big[
   6 p^4 \bigg(\omega _n^2 \big(-939 y^4+11856 \sqrt{14} y^3 z \nonumber \\
   &-&
   7826 y^2 z^2+7784 \sqrt{14} y z^3+4606 z^4\big)+5 \mu ^2 \big(4643 y^4+4080 \sqrt{14} y^3 z+17178 y^2 z^2+1176 \sqrt{14} y z^3 \nonumber \\
   &+& 5586 z^4\big)\bigg)
   +
   80 \mu  p^2 \left(2178 y^4+1089 \sqrt{14} y^3 z+2065 y^2 z^2+1239 \sqrt{14} y z^3-588 z^4\right) \left(\mu ^2+\omega _n^2\right) \\
   &+&
   7 \big(10689 y^4
   +
   1584 \sqrt{14} y^3 z+15946 y^2 z^2-504 \sqrt{14} y z^3+12054 z^4\big) \left(\mu ^2+\omega _n^2\right){}^2+p^8 \big(1527 y^4 \nonumber \\
   &+&
   16272 \sqrt{14} y^3 z+85358 y^2 z^2
   +
   1848 \sqrt{14} y z^3+6762 z^4\big)+16 \mu  p^6 (2466 y^4+4401 \sqrt{14} y^3 z \nonumber \\
   &+&
   23989 y^2 z^2 + 2709 \sqrt{14} y z^3-4704 z^4)
   \big] \nonumber
   \:,
  \end{eqnarray}
  \begin{eqnarray}
   F_4^2(\Delta_{\left|2,2\right>}) &=& T \sum_n \int_0^\Lambda dp \frac{-p^2}{385875 \pi ^2 \left(2 \omega _n^2 \left(\mu ^2+p^4\right)+\omega _n^4+\left(p^4-\mu ^2\right)^2\right){}^2} \big[
   -2 p^4 \bigg(3 \omega _n^2 \big(19 y^4+9784 \sqrt{14} y^3 z\nonumber \\
   &+&
   9576 y^2 z^2+6356 \sqrt{14} y z^3-686 z^4\big)+\mu ^2 (66643 y^4+67288 \sqrt{14} y^3 z+210392 y^2 z^2+24612 \sqrt{14} y z^3 \nonumber  \\
   &+&
   43218 z^4 )\bigg)
   - 8 \mu  p^2 \left(21578 y^4+11003 \sqrt{14} y^3 z+26222 y^2 z^2+7812 \sqrt{14} y z^3-2352 z^4\right) \left(\mu ^2+\omega _n^2\right) \nonumber \\
   &-&
   7 \big(10673 y^4 +
   1568 \sqrt{14} y^3 z+16562 y^2 z^2-168 \sqrt{14} y z^3+4998 z^4\big) \left(\mu ^2+\omega _n^2\right){}^2-p^8 \big(4683 y^4 \\
   &+&
   10608 \sqrt{14} y^3 z + 110642 y^2 z^2+12432 \sqrt{14} y z^3-3822 z^4\big)-8 \mu  p^6 (4926 y^4+8841 \sqrt{14} y^3 z \nonumber \\
   &+&
   47894 y^2 z^2+4284 \sqrt{14} y z^3+1176 z^4)
   \big] \nonumber
   \:,
   \end{eqnarray}
  \begin{eqnarray}
   F_4^2(\Delta_{\left|2,1\right>}) &=& T \sum_n \int_0^\Lambda dp \frac{p^2}{771750 \pi ^2 \left(2 \omega _n^2 \left(\mu ^2+p^4\right)+\omega _n^4+\left(p^4-\mu ^2\right)^2\right){}^2} \big[
   2 p^4 \bigg(\mu ^2 \big(101198 y^4+98048 \sqrt{14} y^3 z \nonumber \\
   &+&
   329602 y^2 z^2+45192 \sqrt{14} y z^3+76293 z^4\big)-3 \omega _n^2 (5698 y^4-21632 \sqrt{14} y^3 z+4662 y^2 z^2 \nonumber \\
   &-&11368 \sqrt{14} y z^3+1323 z^4)\bigg)
   +
   8 \mu  p^2 (25576 y^4+16936 \sqrt{14} y^3 z+49504 y^2 z^2+14049 \sqrt{14} y z^3 \\
   &-&
   3969 z^4) \left(\mu ^2+\omega _n^2\right)+7 \big(11926 y^4
   +
   2656 \sqrt{14} y^3 z+33964 y^2 z^2-336 \sqrt{14} y z^3+9261 z^4\big) \nonumber \\
   &\times&
   \left(\mu ^2+\omega _n^2\right){}^2
   +
   p^8 \big(27354 y^4+20064 \sqrt{14} y^3 z
   + 138376 y^2 z^2+8736 \sqrt{14} y z^3+8379 z^4\big) \nonumber \\
   &+&
   8 \mu  p^6 \left(13416 y^4+13056 \sqrt{14} y^3 z+66304 y^2 z^2+7119 \sqrt{14} y z^3+441 z^4\right)
   \big] \nonumber
   \:.
  \end{eqnarray}

 %\end{widetext}
 %
 For the sextic order, we choose the real states $\frac{1}{\sqrt{2}}(\left| 2,+2\right> +\left| 2,-2 \right>)$ and $\left|2,0\right>$, and find the following matching conditions:
 \begin{eqnarray}
  F_6^2(\Delta_1=1,\Delta_2=0) &=& v_1 , \\
  F_6^2(\Delta_1=0,\Delta_2=1) &=& v_1 + \frac{4}{3} v_2 ,
 \end{eqnarray}
with
%
%\begin{widetext}
%
\begin{eqnarray}
 F_6^2(\Delta_1=1,\Delta_2=0) &=& T \sum_n \int_0^\Lambda dp \frac{-p^2}{17381739375 \pi ^2 \left(2 \omega _n^2 \left(\mu ^2+p^4\right)+\omega _n^4+\left(p^4-\mu ^2\right)^2\right){}^3}
 \big[
 3 p^8 \bigg(65 \omega _n^2 \big(1322251 y^6 \nonumber \\
 &+&
 1818672 \sqrt{14} y^5 z+24967565 y^4 z^2+98280 \sqrt{14} y^3 z^3+23374470 y^2 z^4-1353576 \sqrt{14} y z^5+2302902 z^6\big) \nonumber \\
 &+&
 \mu ^2 \big(377751751 y^6+738633264 \sqrt{14} y^5 z+6372678585 y^4 z^2+1772998920 \sqrt{14} y^3 z^3 \nonumber \\
 &+&
 3520792590 y^2 z^4-207035976 \sqrt{14} y z^5+865354014 z^6\big)\bigg)+24 \mu  p^6 \bigg(13 \omega _n^2 \big(3459534 y^6 \nonumber \\
 &+&
 8171271 \sqrt{14} y^5 z+46236715 y^4 z^2+8805930 \sqrt{14} y^3 z^3+35585760 y^2 z^4+5678316 \sqrt{14} y z^5 \nonumber \\
 &-&
 9390654 z^6\big)+15 \mu ^2 \big(7992198 y^6+10796899 \sqrt{14} y^5 z+70702135 y^4 z^2+18477410 \sqrt{14} y^3 z^3 \nonumber \\
 &+&
 31885280 y^2 z^4+8335292 \sqrt{14} y z^5-11292246 z^6\big)\bigg)+195 p^4 \left(\mu ^2+\omega _n^2\right) \bigg(11 \omega _n^2 \big(28165 y^6  \\
 &+&
 518688 \sqrt{14} y^5 z+36925 y^4 z^2+795480 \sqrt{14} y^3 z^3-186690 y^2 z^4+311640 \sqrt{14} y z^5+228438 z^6\big) \nonumber \\
 &+&
 \mu ^2 \big(19893187 y^6+18064800 \sqrt{14} y^5 z+92444835 y^4 z^2+20433000 \sqrt{14} y^3 z^3+96783330 y^2 z^4 \nonumber \\
 &-&
 1007832 \sqrt{14} y z^5+23677290 z^6\big)\bigg)+5148 \mu  p^2 \big(512558 y^6+293047 \sqrt{14} y^5 z+1153355 y^4 z^2 \nonumber \\
 &+&
 479010 \sqrt{14} y^3 z^3+337120 y^2 z^4+377692 \sqrt{14} y z^5-187278 z^6\big) \left(\mu ^2+\omega _n^2\right){}^2+715 \big(1009801 y^6 \nonumber \\
 &+&
 244656 \sqrt{14} y^5 z+2270835 y^4 z^2+194040 \sqrt{14} y^3 z^3+2438730 y^2 z^4-95256 \sqrt{14} y z^5+1352106 z^6\big)  \nonumber \\
 &\times&
 \left(\mu ^2+\omega _n^2\right){}^3+p^{12} \big(27974687 y^6+55081728 \sqrt{14} y^5 z+1162083195 y^4 z^2+399538440 \sqrt{14} y^3 z^3 \nonumber \\
 &+&
 162521730 y^2 z^4-104354712 \sqrt{14} y z^5+141674778 z^6\big)+36 \mu  p^{10} \big(6703346 y^6+16793049 \sqrt{14} y^5 z\nonumber \\
 &+&
 206611685 y^4 z^2+60932270 \sqrt{14} y^3 z^3+74879840 y^2 z^4+6595204 \sqrt{14} y z^5-29217426 z^6\big)
 \big] \nonumber
\end{eqnarray}
and
\begin{eqnarray}
 F_6^2(\Delta_1=0,\Delta_2=1) &=& T \sum_n \int_0^\Lambda dp
 \frac{-p^2}{17381739375 \pi ^2 \left(\left(\mu ^2+\omega _n^2\right){}^2-2 p^4 \left(\mu ^2-\omega _n^2\right)+p^8\right){}^3}
 \big[
 3 p^8 \bigg(117 y^6 \big(4602891 \mu ^2 \nonumber \\
 &+&
 574015 \omega _n^2\big)+144 \sqrt{14} y^5 z \left(5708537 \mu ^2+1092065 \omega _n^2\right)+1365 y^4 z^2 \left(4847253 \mu ^2+1309465 \omega _n^2\right) \nonumber \\
 &+&
 2520 \sqrt{14} y^3 z^3 \left(692527 \mu ^2+8515 \omega _n^2\right)+490 y^2 z^4 \left(7566047 \mu ^2+2838355 \omega _n^2\right)-2352 \sqrt{14} y z^5 \nonumber \\
 & \times &
 \big(114811 \mu ^2+41015 \omega _n^2\big)+686 z^6 \left(1516423 \mu ^2+266435 \omega _n^2\right)\bigg)-8 \mu  p^6 \bigg(-54 y^6 \big(7616157 \mu ^2 \nonumber \\
 &+&
 2579083 \omega _n^2\big)-459 \sqrt{14} y^5 z \left(1107053 \mu ^2+736827 \omega _n^2\right)-315 y^4 z^2 \left(10222497 \mu ^2+6302023 \omega _n^2\right) \nonumber \\
 &-&
 1890 \sqrt{14} y^3 z^3 \left(462369 \mu ^2+186511 \omega _n^2\right)-8820 y^2 z^4 \left(151831 \mu ^2+141609 \omega _n^2\right)-882 \sqrt{14} y z^5 \nonumber \\
 & \times &
 \big(461771 \mu ^2+268749 \omega _n^2\big)+686 z^6 \left(875293 \mu ^2+605787 \omega _n^2\right)\bigg)+195 p^4 \left(\mu ^2+\omega _n^2\right) \bigg(27 y^6 \\
 &\times &
 \left(754569 \mu ^2-2651 \omega _n^2\right)+576 \sqrt{14} y^5 z \left(32194 \mu ^2+10659 \omega _n^2\right)+105 y^4 z^2 \left(921243 \mu ^2+9911 \omega _n^2\right) \nonumber \\
 &+&
 2520 \sqrt{14} y^3 z^3 \left(8207 \mu ^2+3619 \omega _n^2\right)+490 y^2 z^4 \left(203135 \mu ^2-4301 \omega _n^2\right)-2352 \sqrt{14} y z^5 \big(1189 \mu ^2 \nonumber \\
 &-&
 1111 \omega _n^2\big)+686 z^6 \left(39089 \mu ^2+5269 \omega _n^2\right)\bigg)-1716 \mu  p^2 \big(-1512378 y^6-906237 \sqrt{14} y^5 z \nonumber \\
 &-&
 3801105 y^4 z^2-1409310 \sqrt{14} y^3 z^3-792820 y^2 z^4-1223922 \sqrt{14} y z^5+811538 z^6\big) \left(\mu ^2+\omega _n^2\right){}^2 \nonumber \\
 &+&
 715 \big(985257 y^6+272592 \sqrt{14} y^5 z+2180115 y^4 z^2+264600 \sqrt{14} y^3 z^3+2450490 y^2 z^4 \nonumber  \\
 &-&
 190512 \sqrt{14} y z^5+1452262 z^6\big) \left(\mu ^2+\omega _n^2\right){}^3+5 p^{12} \big(11635299 y^6+24002784 \sqrt{14} y^5 z \nonumber \\
 &+&
 277481295 y^4 z^2+78238440 \sqrt{14} y^3 z^3+32570790 y^2 z^4-22431024 \sqrt{14} y z^5+32344214 z^6\big) \nonumber \\
 &-&
 12 \mu  p^{10} \big(-38287782 y^6-67827123 \sqrt{14} y^5 z-673045695 y^4 z^2-180862290 \sqrt{14} y^3 z^3 \nonumber \\
 &-&
 183172780 y^2 z^4-25635918 \sqrt{14} y z^5+101429902 z^6\big)
 \big] \nonumber \:,
\end{eqnarray}
which yields
 \begin{eqnarray}
  v_1 &=&T \sum_n \int_0^\Lambda dp \frac{-p^2}{17381739375 \pi ^2 \left(2 \omega _n^2 \left(\mu ^2+p^4\right)+\omega _n^4+\left(p^4-\mu ^2\right)^2\right)^3}
  \big[
  3 p^8 \bigg(65 \omega _n^2 \big(1322251 y^6+1818672 \sqrt{14} y^5 z \nonumber \\
  &+&
  24967565 y^4 z^2+98280 \sqrt{14} y^3 z^3+23374470 y^2 z^4-1353576 \sqrt{14} y z^5+2302902 z^6\big)+\mu ^2 \big(377751751 y^6 \nonumber \\
  &+&
  738633264 \sqrt{14} y^5 z+6372678585 y^4 z^2+1772998920 \sqrt{14} y^3 z^3+3520792590 y^2 z^4-207035976 \sqrt{14} y z^5 \nonumber \\
  &+&
  865354014 z^6\big)\bigg)
  +24 \mu  p^6 \bigg(13 \omega _n^2 \big(3459534 y^6+8171271 \sqrt{14} y^5 z+46236715 y^4 z^2+8805930 \sqrt{14} y^3 z^3 \nonumber \\
  &+&
  35585760 y^2 z^4+5678316 \sqrt{14} y z^5-9390654 z^6\big)+15 \mu ^2 \big(7992198 y^6+10796899 \sqrt{14} y^5 z+70702135 y^4 z^2 \nonumber \\
  &+&
  18477410 \sqrt{14} y^3 z^3+31885280 y^2 z^4+8335292 \sqrt{14} y z^5-11292246 z^6\big)\bigg)+195 p^4 \big(\mu ^2+\omega _n^2\big) \bigg(11 \omega _n^2 \big(28165 y^6 \\
  &+&
  518688 \sqrt{14} y^5 z+36925 y^4 z^2+795480 \sqrt{14} y^3 z^3-186690 y^2 z^4+311640 \sqrt{14} y z^5+228438 z^6\big)+\mu ^2 \big(19893187 y^6  \nonumber \\
  &+&
  18064800 \sqrt{14} y^5 z+92444835 y^4 z^2+20433000 \sqrt{14} y^3 z^3+96783330 y^2 z^4-1007832 \sqrt{14} y z^5+23677290 z^6\big)\bigg) \nonumber \\
  &+&
  5148 \mu  p^2 \big(512558 y^6+293047 \sqrt{14} y^5 z+1153355 y^4 z^2+479010 \sqrt{14} y^3 z^3+337120 y^2 z^4+377692 \sqrt{14} y z^5-187278 z^6\big) \nonumber \\
  &\times&
  \left(\mu ^2+\omega _n^2\right){}^2+715 \big(1009801 y^6+244656 \sqrt{14} y^5 z+2270835 y^4 z^2+194040 \sqrt{14} y^3 z^3+2438730 y^2 z^4-95256 \sqrt{14} y z^5  \nonumber \\
  &+&
  1352106 z^6\big) \left(\mu ^2+\omega _n^2\right){}^3+p^{12} \big(27974687 y^6+55081728 \sqrt{14} y^5 z+1162083195 y^4 z^2+399538440 \sqrt{14} y^3 z^3  \nonumber \\
  &+&
  162521730 y^2 z^4-104354712 \sqrt{14} y z^5+141674778 z^6\big)+36 \mu  p^{10} \big(6703346 y^6+16793049 \sqrt{14} y^5 z+206611685 y^4 z^2 \nonumber \\
  &+&
  60932270 \sqrt{14} y^3 z^3+74879840 y^2 z^4+6595204 \sqrt{14} y z^5-29217426 z^6\big)
  \big] \nonumber
 \end{eqnarray}
and
\begin{eqnarray}
 v_2 &=& T \sum_n \int_0^\Lambda dp
 \frac{-p^2}{5793913125 \pi ^2 \left(2 \omega _n^2 \left(\mu ^2+p^4\right)+\omega _n^4+\left(p^4-\mu ^2\right)^2\right)^3}
 \big[
 3 p^8 \bigg(\mu ^2 \big(40196624 y^6+20849016 \sqrt{14} y^5 z \nonumber \\
 &+&
 60955440 y^4 z^2-6957720 \sqrt{14} y^3 z^3+46642610 y^2 z^4-15749874 \sqrt{14} y z^5+43728041 z^6\big)-65 \omega _n^2 \big(72256 y^6 \nonumber \\
 &-&
 150168 \sqrt{14} y^5 z-632800 y^4 z^2-57960 \sqrt{14} y^3 z^3+494410 y^2 z^4+32634 \sqrt{14} y z^5-127253 z^6\big)\bigg)
 \nonumber \\
 &+&4 \mu  p^6 \bigg(39 \omega _n^2 \big(55752 y^6+250308 \sqrt{14} y^5 z+2332120 y^4 z^2+116340 \sqrt{14} y^3 z^3-1780170 y^2 z^4+199773 \sqrt{14} y z^5 \nonumber \\
 &-&
 632492 z^6\big)+\mu ^2 \big(25811784 y^6+11138436 \sqrt{14} y^5 z+19245240 y^4 z^2+21196980 \sqrt{14} y^3 z^3-47844090 y^2 z^4 \\
 &+&
 16096941 \sqrt{14} y z^5-46149964 z^6\big)\bigg)+195 p^4 \left(\mu ^2+\omega _n^2\right) \bigg(4 y^6 \big(30011 \mu ^2-23837 \omega _n^2\big)+72 \sqrt{14} y^5 z \left(1663 \mu ^2+1507 \omega _n^2\right) \nonumber  \\
 &+&
 140 y^4 z^2 \left(7653 \mu ^2+1133 \omega _n^2\right)+1680 \sqrt{14} y^3 z^3 \left(37 \mu ^2+55 \omega _n^2\right)+245 y^2 z^4 \left(2809 \mu ^2-55 \omega _n^2\right) \nonumber  \\
 &-&
 2646 \sqrt{14} y z^5 \left(169 \mu ^2+77 \omega _n^2\right)+343 z^6 \left(2287 \mu ^2+803 \omega _n^2\right)\bigg)
 -858 \mu  p^2 \big(12648 y^6-13548 \sqrt{14} y^5 z-170520 y^4 z^2 \nonumber  \\
 &+&
 13860 \sqrt{14} y^3 z^3+109270 y^2 z^4-45423 \sqrt{14} y z^5+124852 z^6\big) \left(\mu ^2+\omega _n^2\right){}^2-715 \left(2 y^2+7 z^2\right) \big(3068 y^4-3492 \sqrt{14} y^3 z \nonumber \\
 &+&
 602 y^2 z^2+3402 \sqrt{14} y z^3-3577 z^4\big) \left(\mu ^2+\omega _n^2\right){}^3
 +p^{12} \big(7550452 y^6+16233048 \sqrt{14} y^5 z+56330820 y^4 z^2 \nonumber \\
 &-&
 2086560 \sqrt{14} y^3 z^3+83055 y^2 z^4-1950102 \sqrt{14} y z^5+5011573 z^6\big)
 +6 \mu  p^{10} \big(9088872 y^6+8723988 \sqrt{14} y^5 z \nonumber \\
 &+&
 26605320 y^4 z^2-967260 \sqrt{14} y^3 z^3-20733370 y^2 z^4+2925153 \sqrt{14} y z^5-6888812 z^6\big)
 \big] \nonumber
 \:.
\end{eqnarray}

 \subsection{$j=3$}
Using $c_\alpha=c_2/7$ and $c_\beta=c_3/7$, we find the following matching conditions:
 \begin{eqnarray}
  F_4^3(\Delta_{\left|3,3\right>}) &=& c_1 + 9 c_2 \:, \\
  F_4^3(\Delta_{\left|3,0\right>}) &=& c_1 + c_\alpha +\frac{4}{3} c_\beta \:, \\
  F_4^3(\Delta_{\frac{1}{\sqrt{2}}(\left|3,+3\right>+\left| 3,-3\right>)}) &=& c_1 + c_\alpha + \frac{25}{12} c_\beta \:, \\
  F_4^3(\Delta_{\left| 3,+2\right>}) &=& c_1 + 4 c_2 \:.
 \end{eqnarray}
The functions are given by:
\begin{equation}
\scalemath{0.84}{ F_4^3(\Delta_{\left|3,3\right>}) = T \sum_n \int_0^\Lambda dp
 \frac{p^2[17017 \left(\mu ^2+\omega _n^2\right){}^2+4576 p^4 \left(13 \mu ^2+4 \omega _n^2\right)-50622 \mu  p^2 \left(\mu ^2+\omega _n^2\right)+6823 p^8-32370 \mu  p^6]}{60060 \pi ^2 \left(2 \omega _n^2 \left(\mu ^2+p^4\right)+\omega _n^4+\left(p^4-\mu ^2\right)^2\right){}^2}
 \:,}
\end{equation}
 \begin{equation}
  \scalemath{0.84}{F_4^3(\Delta_{\left|3,0\right>}) = T \sum_n \int_0^\Lambda dp
  \frac{ 2  p^2[23023 \left(\mu ^2+\omega _n^2\right){}^2+286 p^4 \left(285 \mu ^2+71 \omega _n^2\right)-68640 \mu  p^2 \left(\mu ^2+\omega _n^2\right)+8459 p^8-43368 \mu  p^6]}{375375 \pi ^2 \left(2 \omega _n^2 \left(\mu ^2+p^4\right)+\omega _n^4+\left(p^4-\mu ^2\right)^2\right){}^2}
  \:,}
 \end{equation}
 \begin{equation}
 \scalemath{0.84}{
  F_4^3(\Delta_{\frac{1}{\sqrt{2}}(\left|3,+3\right>+\left| 3,-3\right>)})
  =
  T \sum_n \int_0^\Lambda dp
  \frac{p^2 [17017 \left(\mu ^2+\omega _n^2\right){}^2+572 p^4 \left(105 \mu ^2+31 \omega _n^2\right)-50622 \mu  p^2 \left(\mu ^2+\omega _n^2\right)+6611 p^8-32370 \mu  p^6]}{120120 \pi ^2  \left(2 \omega _n^2 \left(\mu ^2+p^4\right)+\omega _n^4+\left(p^4-\mu ^2\right)^2\right){}^2}
  \:,}
 \end{equation}
 \begin{equation}
 \scalemath{0.84}{ F_4^3(\Delta_{\left| 3,+2\right>})
  =
  T \sum_n \int_0^\Lambda dp
  \frac{p^2 [7007 \left(\mu ^2+\omega _n^2\right){}^2+572 p^4 \left(43 \mu ^2+9 \omega _n^2\right)-20878 \mu  p^2 \left(\mu ^2+\omega _n^2\right)+2533 p^8-13026 \mu  p^6]}{45045 \pi ^2 \left(2 \omega _n^2 \left(\mu ^2+p^4\right)+\omega _n^4+\left(p^4-\mu ^2\right)^2\right){}^2}
  \:.}
 \end{equation}

 \end{widetext}

\end{document}